\documentstyle[aaspp4]{article}
\slugcomment{Accepted for publication in \it The Astronomical Journal\rm}

\lefthead{Gibson et~al.}
\righthead{Metal Abundances in the Magellanic Stream}

\begin{document}

\title{Metal Abundances in the Magellanic Stream\altaffilmark{1}}

\author{Brad K. Gibson\altaffilmark{2}, 
	Mark L. Giroux\altaffilmark{2},
	Steven V. Penton\altaffilmark{2}, \break
	Mary E. Putman\altaffilmark{3},
	John T. Stocke\altaffilmark{2} and
        J. Michael Shull\altaffilmark{2,4}}
\altaffiltext{1}{Based upon observations with the NASA/ESA Hubble Space
		 Telescope, obtained at the Space Telescope Science 
		 Institute, which is operated by the Association of
		 Universities for Research in Astronomy, Inc., under
		 NASA contract NAS5-26555.}
\altaffiltext{2}{Center for Astrophysics \& Space Astronomy,
                 Department of Astrophysical \& Planetary Sciences, 
		 Campus Box 389, University of Colorado, Boulder, CO 80309-0389}
\altaffiltext{3}{Research School of Astronomy \& Astrophysics,
                 Australian National University, Weston Creek Post Office,
                 Weston, ACT, Australia 2611}
\altaffiltext{4}{Also at JILA, University of Colorado and National
		 Institute of Standards and Technology}

\def\spose#1{\hbox to 0pt{#1\hss}}
\def\simlt{\mathrel{\spose{\lower 3pt\hbox{$\mathchar"218$}}
     \raise 2.0pt\hbox{$\mathchar"13C$}}}
\def\simgt{\mathrel{\spose{\lower 3pt\hbox{$\mathchar"218$}}
     \raise 2.0pt\hbox{$\mathchar"13E$}}}
\def\kms{{\rm km\,s$^{-1}$}}
\def\eg{{\rm e.g.}}
\def\ie{{\rm i.e.}}
\def\etal{{\rm et~al.}}
\def\ho{{\rm H$_\circ$}}
\def\h0{{\rm H$_\circ$}}
\def\hounits{{\rm km\,s$^{-1}$\,Mpc$^{-1}$}}
\def\tf{{\rm Tully-Fisher }}
\def\ct{{\rm Cal\'{a}n-Tololo }}
\def\bmv{\hbox{\it B--V\/}}
\def\vmi{\hbox{\it V--I\/}}

\begin{abstract}

We report on the first metallicity determination for gas in
the Magellanic Stream, using archival \it HST \rm GHRS data for the background
targets Fairall~9, III~Zw~2, and NGC~7469.  For Fairall~9, using
two subsequent \it HST \rm revisits and new Parkes Multibeam Narrowband 
observations, we have unequivocally detected the MS\,I \ion{H}{1}
component of the Stream (near its head) 
in \ion{S}{2}\,$\lambda\lambda$1250,1253
yielding
a metallicity of [\ion{S}{2}/H]$=-0.55\pm 0.06 ({\rm r})
^{+0.17}_{-0.21} ({\rm s})$,
consistent with either an SMC or LMC origin
and with the earlier upper limit set by Lu, Savage \& Sembach (1994).
We also detect
the saturated \ion{Si}{2}\,$\lambda$1260 line,
but set only a lower limit of [\ion{Si}{2}/H]$\simgt$$-$1.5.  We present
two \it HST \rm serendipitous detections of the Stream, seen in 
\ion{Mg}{2}\,$\lambda\lambda$2796,2803
absorption with column densities of (0.5$-$1)$\times 10^{13}$\,cm$^{-2}$
toward the Seyfert galaxies
III~Zw~2 and NGC~7469.  These latter sightlines probe
gas near the tip of the Stream ($\sim$15$^\circ$ from the peak of the 
MS\,V \ion{H}{1} component and $\sim$80$^\circ$ down-Stream of Fairall~9).
In the case of III~Zw~2, the lack of an accurate \ion{H}{1} column density
determination and the uncertain \ion{Mg}{3} ionization correction severely
limit the degree to which we can constrain [Mg/H]; 
we found a lower limit of [\ion{Mg}{2}/\ion{H}{1}]$\simgt$$-$1.3 
for this sightline.  For NGC~7469, an
accurate \ion{H}{1} column density determination exists, but the extant FOS
spectrum limits our ability to constrain the \ion{Mg}{2} column density, and we
conclude that
[\ion{Mg}{2}/\ion{H}{1}]$\simgt$$-$1.5 for this sightline.
Ionization corrections associated with \ion{Mg}{3}
and \ion{H}{2} suggest that the corresponding
[Mg/H] may range lower by $\sim$0.3$-$1.0\,dex.
However, an upward revision of $\sim$0.5$-$1.0\,dex would be expected under 
the assumption that the Stream exhibits a dust depletion pattern similar to that
seen in both the Large and Small Magellanic Clouds.
While our abundance analysis 
allows us to rule out a primordial origin for the Stream, the remaining
systematic uncertainties in the \ion{H}{1} column density along the lines of
sight makes it difficult to differentiate between an LMC versus SMC origin.
\end{abstract}


\section{Introduction}
\label{intro}

Subtending $\sim$100$^\circ$ from the Magellanic Clouds through the south
Galactic pole and beyond, the Magellanic Stream is the most striking
extragalactic feature in the neutral hydrogen sky (Wannier \& Wrixon 1972).
Despite its prominence and apparent association with the Clouds, the Stream's
origin remains somewhat controversial. 
Tidally-disrupted (Putman \etal\ 1998)
or ram pressure-stripped (Moore \& Davis 1994) gas from the LMC or SMC remain
the most likely scenarios, although ram pressure-stripped gas from the
inter-Cloud region has also been suggested (Mathewson \etal\ 1987).  

A related question pertains to the still uncertain relationship between the 
Stream
and the general population of Galactic High-Velocity Clouds
(HVCs).\footnote{The Magellanic
Stream corresponds to HVC~\#493 in the Wakker \& van Woerden
(1991) catalog of High-Velocity Clouds.  In our study, we use the
classical definition of the Stream - \ie, the stream of gas 
\it trailing \rm the Magellanic Clouds.  Other HVCs \it leading \rm
the Clouds have been ascribed
a similar origin to that of the Stream (Lu \etal\ 1998; Putman \etal\ 1998;
Sahu 1998), but for our purposes the conventional definition is assumed
throughout.}
As reviewed by Wakker \& van Woerden (1997, \S~6.3),
several models ascribe 
a Magellanic Cloud origin to
some of the most prominent
HVC complexes (\eg, Complex~C, VHVCs, Population~EP),
similar to that of the Stream.

The chemical composition of gas comprising the Stream and HVCs has
long been recognized as a key discriminant between competing models for their
respective origins.  Unfortunately, the dearth of suitable background probes
against which to determine abundances of intervening HVC gas has made progress
in this area difficult.  To date, accurate metallicities have been derived for
only a few HVCs, including
[\ion{S}{2}/\ion{H}{1}]$=-0.60\pm 0.13$, using the background probe NGC~3783,
for HVC~287+22+240 (Lu \etal\ 1998), and
[S/H]$=-1.03\pm 0.10$, using Mrk~290, for Complex~C (Wakker \etal\ 1999).
A detailed study of the metallicity distribution within Complex~C,
using five different background probes, will
be reported in a future paper (Gibson \etal\ 2000).

A comparison of the metallicity of the Stream with these and other
HVCs is hampered by this same lack of suitable background probes.
All efforts to
date have concentrated upon the background Seyfert galaxy Fairall~9,
which intersects the MS~I concentration of the Stream
(see Figure~1 of Mathewson 1985 for nomenclature), $\sim$10$^\circ$
down-Stream from the SMC.  Songaila (1981) detected the Stream in \ion{Ca}{2}
optical absorption, which when
combined with the lack of a detection at \ion{Na}{1}~D, and
subsequent ionization equilibrium arguments, led to the loose constraint
that $-2\simlt [{\rm Ca/H}]\simlt +0.25$.  This limit was not stringent
enough to differentiate between a primordial or Magellanic Cloud origin for the
Stream.

More recently, Lu, Savage \& Sembach (1994) examined
two Stream clouds 
along the Fairall~9 sightline
(one at $+210$\,km\,s$^{-1}$, the other at $+170$\,km\,s$^{-1}$).  They
derived \ion{S}{2} $\lambda\lambda$1250,1253 and 
\ion{Si}{2} $\lambda\lambda$1260,1526 abundance limits of 
[\ion{S}{2}/\ion{H}{1}]$\simlt -0.52$
and [\ion{Si}{2}/\ion{H}{1}]$\simgt 
-1.15$, for the $+210$\,km\,s$^{-1}$ cloud, and
[\ion{S}{2}/\ion{H}{1}]$\simlt 
-0.05$ and [\ion{Si}{2}/\ion{H}{1}]$\simgt -0.70$, for the $+170$\,km\,s$^{-1}$
cloud.  Under the assumption that the two clouds have the same metallicity, 
these constraints, for the full sightline, reduce to 
[\ion{S}{2}/\ion{H}{1}]$\simlt -0.52$ and
[\ion{Si}{2}/\ion{H}{1}]$\simgt -0.70$.  The low
signal-to-noise ratio S/N in the vicinity of the
sulfur lines only allowed Lu et~al. to set
upper limits on [S/H]; the
saturated silicon lines allowed for only lower limits on 
[\ion{Si}{2}/\ion{H}{1}].

As part of our ongoing \it HST \rm Guest Observer
program on the origin and physical
conditions in the local Ly$\alpha$ forest 
(cf. Penton, Stocke, \& Shull 2000; Penton, Shull,
\& Stocke 2000), we 
revisited Fairall~9 with a similar GHRS set-up to that employed by Lu
\etal\ (1994).  Supplemented with the archival Lu \etal\ dataset, this has
led to a four-fold increase in the effective integration time spent on this one
target; the resultant increase in S/N has allowed us, for the first time, to
derive an unequivocal metallicity determination for the Magellanic Stream.

In \S~2.1, we describe both the GHRS and neutral hydrogen data
employed in our analysis of the Fairall~9 sightline.  The analysis of the
Stream
absorption features seen in this dataset, and the corresponding implications
for 
[\ion{S}{2}/\ion{H}{1}] and [\ion{Si}{2}/\ion{H}{1}], 
are described in \S~3.1.  We report
serendipitous detections of the Stream (but $\sim$80$^\circ$ down-Stream of
Fairall~9, near the tip and adjacent to the MS\,V concentration) 
in \S~2.2, 3.2, 2.3, and 3.3.
These detections of the Stream, seen in \ion{Mg}{2}
absorption in the spectra of III~Zw~2 and NGC~7469, 
represent the only other probes, to date, toward
which the Stream has been detected.  
We discuss the implications of our detections
in \S~4.
Our results are summarized in \S~5.

\section{Data}
\label{data}

\subsection{Fairall~9}
\label{data_f9}

The first three entries of Table~1 list the \it HST \rm GHRS data
employed in our analysis of the Fairall~9 sightline.  A single, merged,
spectrum was created by first scaling the flux levels of the two 1996 exposures
-- \tt z3e70404m \rm and \tt z3e70406t \rm -- to be consistent with the initial
Lu \etal\ (1994) exposure (\tt z26o0208n\rm),\footnote{This scaling was
applied linearly across the region overlapping that of 
the initial exposure.} and then
averaging the three, weighting by the flux uncertainties.  A detailed
description of the spectrum preparation can be found in Penton,
Stocke \& Shull (2000).
The effective integration time of the merged Fairall~9 spectrum totals
$\sim$7.5\,hrs, with a 3$\sigma$ equivalent width detection limit of 6\,m\AA\
at $\lambda=1250$\,\AA.  The S/N at $\lambda=1250$\,\AA\ is $\sim$30.

\placetable{tbl:obs}

The resulting
merged GHRS spectrum is shown in Figure 1.  The Galactic
and Stream lines of \ion{S}{2} $\lambda\lambda$1250, 1253 and \ion{Si}{2} $\lambda$1260 are
labeled accordingly.  The three Galactic lines shown are redshifted (in the
mean) by 0.028\,\AA, with respect to their rest frame wavelengths (taken from
Verner \etal\ 1994, and listed in Table 3), corresponding to
$\sim$$+$7\,km\,s$^{-1}$ at $\lambda=1250$\,\AA.  In contrast, the Galactic
\ion{H}{1} along this sightline (upper panel of
Figure 4) peaks at $v_{\rm
LSR}\approx -1$\,km\,s$^{-1}$.  This suggests that an \it a posteriori \rm
velocity shift of $\Delta v\approx -8$\,km\,s$^{-1}$ might be appropriate for
our merged GHRS Fairall~9 spectrum,\footnote{Such a shift was applied by 
Penton, Stocke \& Shull (2000) in the Ly$\alpha$ absorber
analysis.} although we have not done so here 
since it has no impact upon the Stream abundance analysis that follows.

\placefigure{fig:F9_GHRS}

The neutral hydrogen column density of the 
Stream in this direction
is N(\ion{H}{1})=$9.35\times 10^{19}$\,cm$^{-2}$.  The two-component 
velocity structure of the high-velocity gas (components centered at
$+$155\,km\,s$^{-1}$ and $+$195\,km\,s$^{-1}$)
has been noted before, although the spectrum shown in Figure~4
is the first to clearly show it.  We will return to a discussion of the
\ion{H}{1} properties of this sightline in \S~3.1.

\subsection{III~Zw~2}
\label{data_mrk1501}

The second Stream background probe discussed here is III~Zw~2.  As evidenced by
its entry in Table~1 (row 4), only a single 25\,min \it HST \rm
GHRS G270M exposure is available.  The corresponding 3$\sigma$ equivalent width
detection limit and S/N, at $\lambda=2800$\,\AA, are 27\,m\AA\ and 8,
respectively.  The spectrum itself is shown in Figure 2,
with Galactic and Stream
\ion{Mg}{2} $\lambda\lambda$2796, 2803 features identified.

\placefigure{fig:IIIZw2_GHRS}

The Galactic \ion{Mg}{2} lines are blueshifted (in the mean) by 0.236 \AA\ (\ie,
$\sim$25\,km\,s$^{-1}$ at $\lambda=2800$\AA), with respect
to their rest frame wavelengths (Pickering, Thorne \& Webb 1998).  This offset
is to be expected, based upon
the \ion{H}{1} distribution along this line of sight, which shows a
two-component structure (Galactic and intermediate-velocity gas) centered on
$v_{\rm LSR}\approx -30$\,km\,s$^{-1}$.
As was the case for our Fairall~9 spectrum (\S~2.1), we have not
applied any \it a posteriori \rm velocity shift to the GHRS data.

\subsection{NGC~7469}
\label{data_ngc7469}

As part of an \it HST \rm FOS snapshot program, NGC~7469 was observed with
the G270H grating for 5\,min during 1996.  The 3$\sigma$ equivalent width
detection limit and S/N at $\lambda=2800$\,\AA~are 162\,m\AA\ and 29,
respectively.  The FOS spectrum is shown in Figure~3,
with Galactic and Stream
\ion{Mg}{2} $\lambda\lambda$2796, 2803 features identified.  An \it a
posteriori \rm velocity shift of $-$20\,km\,s$^{-1}$ has been applied to 
the FOS wavelength calibration, in order to reconcile a systematic offset
between the Galactic and Stream \ion{Mg}{2}\,$\lambda\lambda$2796,2803
and \ion{H}{1} features.  The NRAO 140$^\prime$ \ion{H}{1} spectrum employed in
our analysis (upper panel of Figure~8) was kindly provided
by Ken Sembach prior to publication (Murphy, Sembach \& Lockman 2000).

\placefigure{fig:n7469_Mg_FOS}

\section{Analysis}
\label{analysis}

\subsection{Fairall~9}
\label{analysis_f9}

Figure 4 shows the velocity stack for the Galactic (G) and
Stream (MS) lines 
(\ion{S}{2} $\lambda$1250, \ion{S}{2} $\lambda$1253, and \ion{Si}{2} $\lambda$1260)
detected in our Fairall~9 GHRS spectrum, as well as an \ion{H}{1} spectrum
taken with the Parkes Narrowband Receiver (Haynes \etal\ 1999 - spectrum
kindly provided by Lister Staveley-Smith).
The \ion{S}{2} $\lambda$1259 Stream line is blended with the 
Galactic \ion{Si}{2} $\lambda$1260 line in the lower panel.  In Table
2, we list the centroids of the Stream and Galactic lines
(column 1), velocity range over which equivalent widths were determined
(column 2), and the line equivalent widths (column 3).\footnote{Equivalent
widths were derived by integrating over the line profile; Gaussian 
fits were also undertaken (cf. Penton, Stocke \& Shull 2000), but the
results here are insensitive to the measurement technique employed, and thus we
only report the former.}  The inferred ionic column densities are listed in
columns 4 and 5 -- the former (N$_{\tau=0}$) corresponds to the optically-thin 
assumption, while the latter (N$_{\tau_v}$) corresponds to the \it
apparent optical
depth method \rm of Sembach \& Savage (1992).  
For the Fairall~9 \ion{S}{2} Stream features, N$_{\tau=0}$ and N$_{\tau_v}$
are consistent within the
uncertainties.
Since the \ion{Si}{2}\,$\lambda$1260 lines are clearly saturated, it is not
surprising that N$_{\tau=0}$ and N$_{\tau_v}$ are discrepant.  This saturation
allows us only to set lower limits on the \ion{Si}{2} column densities.
In what follows, we adopt N$_{\tau_v}$ in deriving the Stream's metallicity,
and we make no attempt to treat the two components seen in the 
\ion{H}{1} spectrum separately in our analysis.

\placefigure{fig:F9_stack}

\placetable{tbl:F9_lines}

In the weak-line (optically-thin) limit, 
the column density N$_{\tau=0}$ (in cm$^{-2}$)
and equivalent width of a line
W$_\lambda$ (in m\AA) are related through
\begin{equation}
{\rm N}_{\tau=0} = 1.13\times 10^{17}\,{{{\rm W}_\lambda}\over{f\lambda_0^2}}, 
\label{eq:column}
\end{equation}
\noindent
where $\lambda_0$ (in \AA) is the rest wavelength of the line and $f$ is its
oscillator strength (\eg, Savage \& Sembach 1996; equation 3).  For the lines
considered in this paper, the relevant values of $\lambda_0$ and $f$
are provided in columns 2 and 3 of Table 3.

\placetable{tbl:atomic}

Under the apparent optical depth method of Sembach \& Savage (1992), the
apparent column density N$_{\tau_v}$, in the limit of a finite number of data
points, is instead given by
\begin{equation}
{\rm N}_{\tau_v} = {{3.767\times 10^{14}}\over{f\lambda_0}}\,\sum_{i=1}^n
\ln\biggl[{{I_c(v_i)}\over{I(v_i)}}\biggr]\,{\rm d}v_i,
\label{eq:app_col}
\end{equation}
\noindent
where $I(v_i)$ and $I_c(v_i)$ are the observed and estimated continuum
intensities at velocity $v_i$ (equations A21 and A29 of Sembach \& Savage).
The statistical noise uncertainty associated with the apparent column density
of equation~2 is
\begin{equation}
\sigma_{{\rm N}_{\tau_v}} = {\rm N}_{\tau_v}\,\Biggl(
\sqrt{\sum_{i=1}^n\sigma^2(v_i)\big/I(v_i)^2\,({\rm d}v_i)^2}
\Biggr/\sum_{i=1}^n\ln\biggl[{{I_c(v_i)}\over{I(v_i)}}\biggr]\,{\rm d}v_i
\Biggr),
\label{eq:sig_app_col}
\end{equation}
\noindent
based upon equations~A27 and A30 of Sembach \& Savage.  Uncertainties quoted in
this analysis correspond to those of equation~3; those
associated with continuum placement have not been considered here.  For the
high S/N Fairall~9 data, the uncertainty associated with the
continuum placement is only an additional $\sim$5\% beyond that of
equation~3 (Penton, Stocke \& Shull 2000) and is
neglected here.

The equivalent width ratio of the \ion{S}{2} Stream features (from
Table~2) is 
W$_\lambda$(\ion{S}{2}\,$\lambda$1253)/W$_\lambda$(\ion{S}{2}\,$\lambda$1250)=
1.72$\pm$0.24, further evidence that any saturation effects are mild (the
expected theoretical ratio is 2.0).  Regardless, allowing for these marginal
optical depth effects, the apparent column densities of
the Stream \ion{S}{2} and \ion{Si}{2} features
seen in the Fairall~9 GHRS spectrum are
N$_{\tau_v}$(\ion{S}{2}\,$\lambda$1250)=$(5.39\pm 0.61)\times
10^{14}$\,cm$^{-2}$,
N$_{\tau_v}$(\ion{S}{2}\,$\lambda$1253)=$(4.75\pm 0.41)\times
10^{14}$\,cm$^{-2}$, and
N$_{\tau_v}$(\ion{Si}{2})$\equiv$N(\ion{Si}{2})$>$$9.29\times
10^{13}$\,cm$^{-2}$.
The \ion{S}{2}-weighted average is
N$_{\tau_v}$(\ion{S}{2})$\equiv$N(\ion{S}{2})=
$(4.95\pm 0.34)\times 10^{14}$\,cm$^{-2}$.

Expressing the sulfur abundance in terms of the logarithmic abundance 
A$_{\rm S}$ with respect to that of the Sun A$_{\rm S_\odot}$, we can write
\begin{equation}
[{\rm S/H}]=\log{{{\rm A}_{\rm S}}\over{{\rm A}_{\rm S_\odot}}}=
\log{\rm N(S)}-\log{\rm N(H)}-\log{\rm A_{\rm S_\odot}},
\label{eq:S_metallicity}
\end{equation}
\noindent
since A$_{\rm S}\equiv$N(S)/N(H), by definition.
The nondetection of \ion{S}{1} $\lambda$1262 in our Fairall 9
spectrum (W$_\lambda$ $< 6$ m\AA) implies
N(\ion{S}{1}) $<2.1 \times 10^{14}$\,cm$^{-2}$. 
With
an ionization potential of only 10.36\,eV, 
\ion{S}{1} is easily ionized by the
ambient halo radiation field and likely
makes a negligible contribution to N(S). 
We currently have no observational
limits on the contribution of
\ion{S}{3} to N(S) but, after Wakker \etal\ (1999)
who consider a Complex~C cloud of similar \ion{H}{1} column density,
we assume that
N(\ion{S}{3})$\ll$N(\ion{S}{2}).  
(Also, see our discussion of corrections for \ion{H}{2} below).
With A$_{\rm S_\odot}=(1.862\pm
0.215)\times 10^{-5}$ (column 4 of Table~3), and explicitly
restricting ourselves to \ion{S}{2} and \ion{H}{1} (as opposed to the 
global S and H), equation~4 reduces to
\begin{equation}
[{\rm SII/HI}] = (19.425\pm 0.058) - \log{\rm N(HI)}.
\label{eq:S_F9}
\end{equation}

During a 02/17/99-02/22/99 observing run at Parkes Observatory, the 
Multibeam Narrowband Facility (Haynes \etal\ 1999) was employed to obtain
the Fairall~9 sightline \ion{H}{1} spectrum
shown in the upper panel of Figure~4.  
The quality of this \ion{H}{1} data is a vast
improvement over that available to Lu \etal\ (1994) - \ie, 
the McGee, Newton \& Morton (1983) and Morras (1983) datasets.
The \ion{H}{1} column density for the Stream, along this sightline, is
N(\ion{H}{1})$=(9.35\pm 0.47)\times 10^{19}$\,cm$^{-2}$.

In combination with equation~5, we can use this new determination
of N(\ion{H}{1}) to derive the sulfur abundance for the Stream in the Fairall~9
sightline, resulting in [\ion{S}{2}/\ion{H}{1}]=$-$0.55$\pm$0.06.  If we
had chosen to use N$_{\tau=0}$, as opposed to N$_{\tau_v}$, the derived value
of [\ion{S}{2}/\ion{H}{1}] would have been reduced by only 0.04\,dex.
If 25\% of the total
hydrogen column is in the form of ionized hydrogen, as appears to be
the case for the
Mrk~290 sightline through 
Complex~C (Wakker \etal\ 1999), the derived [\ion{S}{2}/\ion{H}{1}] could
be 0.12\,dex greater than [\ion{S}{2}/H].
This is (perhaps) a conservative overestimate since, in the 
ionized gas, additional sulfur may be in the form of \ion{S}{3}.
 
A further source of systematic error in our estimate of N(H) is far
more challenging to quantify.  
Our \ion{H}{1} column was derived using the
14$^\prime$ beam provided by
the Parkes dish, while the \it HST \rm
absorption measurements toward Fairall~9
sample the gas at sub-arcsecond resolution.
Comparisons have been made of N(\ion{H}{1}) derived from 
21cm mapping with measurements of N(\ion{H}{1}) derived
from Ly$\alpha$ absorption along the lines of sight
to early type stars in the Galactic halo
(Lockman, Hobbs \& Shull 1986;  Savage \etal\ 2000; Wakker \& Savage 2000).
The ratio N(\ion{H}{1})$_{\rm Ly\alpha}$ / N(\ion{H}{1})$_{\rm 21cm}$
is somewhat less than unity, with a dispersion less than $\pm$50\%.
This low dispersion is in apparent
contrast with the results of 21cm surveys of different angular
resolution, which show variations in N(\ion{H}{1}) as large as
a factor of five over arcminute spatial scales (Wakker \& Schwarz 1991).  

A simple, yet elegant, argument put forth by Bart Wakker suggests
that the two observations may not be in contradiction with one another.
Radio surveys with higher angular resolution (1$^\prime$) reveal that
the number of fields with a given N(\ion{H}{1}) is a steep
inverse power law function of N(\ion{H}{1}).  
This implies that higher resolution
probes of a large beam will be more likely to intersect
low N(\ion{H}{1}) sightlines.\footnote{Of the seven sightlines for which
both $<$1$^\prime$ and 10$^\prime$$-$12$^\prime$ \ion{H}{1} data
exists (Table~A1 of Wakker 2000), in only 4/7 of these cases is
N(\ion{H}{1})$_{1^\prime}$$<$N(\ion{H}{1})$_{10^\prime}$.  The ratio of
N(\ion{H}{1})$_{1^\prime}$$/$N(\ion{H}{1})$_{10^\prime}$ for all
seven is 0.90$\pm$0.13, with extrema of 0.75 and 1.24.  This might lead one to
conclude that the importance of the 
predicted ``resolution effect'' has been overestimated, 
but it should be stressed that it is based upon only seven data points.  To
be conservative we have retained the assumed $\pm$50\% uncertainty in
N(\ion{H}{1}).}
In principle, this same
argument could be extended to much smaller scales,
as the gulf between high resolution 1$^\prime$ 21cm maps and the scales
probed by absorption lines remains large.  

To be conservative, we have incorporated
a factor $\pm$0.17\,dex ($\pm$50\%) in our systematic error budget,
to reflect the 
systematic uncertainty in our estimate of N(\ion{H}{1}).  It remains possible
that we have been unfortunate enough to encounter a line of sight
with $\langle$N(\ion{H}{1})$\rangle$ that differs from the 
``pencil beam'' N(\ion{H}{1})
by a factor of five or more.  Higher resolution radio mapping will
better address the likelihood of this possibility.  It is reassuring to note
that, within the measurement errors, the \ion{S}{2} 
and \ion{H}{1} velocity profiles are similar (Figure~4).
This lends some credence to the suggestion that the
\ion{H}{1} revealed by our Parkes
data is representative of the ``pencil beam'' \ion{H}{1} along the Fairall~9
sightline.

Adding this second
factor in quadrature with the systematic uncertainty in converting
\ion{H}{1} to H, we write our
final derived value for the sulfur abundance of the Stream in the
Fairall~9 sightline as
\begin{equation}
[{\rm SII/H}]=-0.55\pm 0.06({\rm r})\,\,^{+0.17}_{-0.21}({\rm s}),
\label{eq:F9_final}
\end{equation}
\noindent
where `r' and `s' correspond to the associated 
random and systematic uncertainties, respectively.
This result can now be compared with that derived in the earlier Lu
\etal\ (1994) study.  Their best \ion{S}{2} constraint was set by their
3$\sigma$ upper limit on the \ion{S}{2}\,$\lambda$1253 absorption,
W$_\lambda\le 87$\,m\AA\ (consistent with our result of W$_\lambda=
62\pm 5$\,m\AA).  Using our \ion{H}{1} column
density for this sightline, as opposed to the Morras (1983) value used by Lu
et~al., results in an upper limit to the sulfur abundance
[\ion{S}{2}/\ion{H}{1}]$\le -0.46$, consistent with
our result of [\ion{S}{2}/\ion{H}{1}]=$-$0.55$\pm$0.06.

For comparison, the gas-phase sulfur abundances of the Magellanic Clouds
are [S/H]$_{\rm LMC}=-0.57\pm 0.10$ and [S/H]$_{\rm SMC}=-0.68\pm 0.16$ (Russell
\& Dopita 1992).  Because of the magnitude of the residual 
uncertainties (particularly those of a systematic nature), 
associating the MS\,I gas with only one of the Clouds remains 
difficult.   Equation~6 has a somewhat ad hoc provision for 
ionization corrections, and no provision for potential dust depletion effects.
Fortunately, there is some confidence that
\ion{S}{2} is effectively free of both of these
complicating effects (Wakker \etal\ 1999).

In analogy to the derivation of equation~6, we can use the
\ion{Si}{2}\,$\lambda$1260
column density (Table 2)
to derive a lower limit on the MS\,I silicon metallicity of
\begin{equation}
[{\rm SiII/H}]>-1.55\pm 0.03({\rm r})\,\,^{+0.17}_{-0.21}({\rm s}).
\label{eq:F9_Si_final}
\end{equation}
\noindent
Because we are limited to the saturated 
\ion{Si}{2}\,$\lambda$1260\, line, we could derive only the
above lower limit with the current data.  Lu
\etal\ (1994) find W$_\lambda = 449\pm 53$\,m\AA\ for the
\ion{Si}{2}\,$\lambda$1526 Stream absorption.
For the total line of sight \ion{H}{1} column density
N(\ion{H}{1})=$9.35\times 10^{19}$\,cm$^{-2}$, this
corresponds to a lower limit
on the silicon abundance of [\ion{Si}{2}/\ion{H}{1}]$\simgt$$-$1.1.  This
particular limit is set by the much lower S/N
\ion{Si}{2}\,$\lambda$1526 line, 
which is not included in our G160M GHRS spectra.  Therefore, we retain the 
more conservative limit set by \ion{Si}{2}\,$\lambda$1260 
(\ie, equation~7).

\subsection{III~Zw~2}
\label{analysis_mrk1501}

Figure~5 shows the velocity profiles of the detected
Galactic (G) and Magellanic Stream (MS)
lines of \ion{Mg}{2}\,$\lambda\lambda$2796,2803\, in our
G270M GHRS spectrum of III~Zw~2.  The centroids, equivalent widths, and
inferred column densities (both optically thin and following the apparent
optical depth methodology described in \S~3.1) are listed in
Table~4.  While our GHRS spectrum is clearly of low
signal-to-noise (S/N=8 at $\lambda=2800$\,\AA), 
both Stream \ion{Mg}{2} lines appear unsaturated, supported by the fact that the
ratio of their equivalent widths,
W$_\lambda$(\ion{Mg}{2}\,$\lambda$2796)/W$_\lambda$(\ion{Mg}{2}\,$\lambda$2803)=
2.11$\pm$0.94, is consistent with the expected theoretical ratio of 2.0.
N$_{\tau=0}$ and N$_{\tau_v}$ agree to within the uncertainties, further
evidence that only a mild degree of saturation is present.

\placefigure{fig:IIIZw2_stack}

\placetable{tbl:IIIZw2_lines}

Employing the apparent optical depth technique, we derive \ion{Mg}{2} column
densities of
N(\ion{Mg}{2} $\lambda$2796)=$(1.08\pm 0.10)\times 10^{13}$\,cm$^{-2}$ and
N(\ion{Mg}{2} $\lambda$2803)=$(0.91\pm 0.14)\times 10^{13}$\,cm$^{-2}$.
We adopt the weighted average in the analysis which follows,
N(\ion{Mg}{2})=$(1.02\pm 0.08)\times 10^{13}$\,cm$^{-2}$.
Analogous to the derivation of equation~5, 
we can use N(\ion{Mg}{2}) along the line of sight to III~Zw~2 to write
\begin{equation}
[{\rm Mg\,II/H\,I}]=(17.429\pm 0.039) - \log{\rm N(H\,I)}.
\label{eq:Mg_mrk1501}
\end{equation}
\noindent
Unlike the case for Fairall~9, we have little in the way of
useful observational constraints on the $\log{\rm N(H\,I)}$ term of equation
8.  The upper panel of Figure~5 shows
the Hanning-smoothed \ion{H}{1} spectrum from Hartmann \& Burton (1997); the
non-detection of \ion{H}{1} and the position and velocity of our \ion{Mg}{2}
detection sets an upper limit to the \ion{H}{1} column density of
$\sim$5$\times$10$^{18}$\,cm$^{-2}$.\footnote{The 5$\sigma$ detection limit for
the Leiden-Dwingeloo Survey (Hartmann \& Burton 1997) is T$_{\rm B}$=0.35\,K
(unsmoothed).  For an HVC with a linewidth of 8\,km\,s$^{-1}$, the 5$\sigma$
\ion{H}{1} column density detection limit corresponds to
5$\times$10$^{18}$\,cm$^{-2}$; for a 20\,km\,s$^{-1}$ linewidth,
N(\ion{H}{1})=5$\times$10$^{18}$\,cm$^{-2}$ corresponds to a 2$\sigma$ detection
limit.}

The MS\,V concentration of the Stream is
centered on $(\ell,b)\approx (92^\circ,
-51^\circ)$, with $v_{\rm LSR}\approx -350$\,km\,s$^{-1}$ (Mathewson 1985).
The upper panel of Figure~6 
shows an $\sim$2000\,deg$^2$ region
encompassing all of MS\,V, including the III~Zw~2 sightline shown in the
lower panel and the neighboring NGC~7469 sightline discussed
in \S~3.3.
The lower panel of shows contours of neutral hydrogen, derived from the
Leiden-Dwingeloo Survey (Hartmann \& Burton 1997), in a
$14^\circ\times 7^\circ$ region centered on the III~Zw~2 sightline.  The
\ion{H}{1}
contour levels are $3\times 10^{18}$\,cm$^{-2}$, with the outermost level being
N(\ion{H}{1})=$1\times 10^{18}$\,cm$^{-2}$.  
Only the velocity range $-400<v_{\rm LSR}<-250$\,km\,s$^{-1}$
was included in this figure.  The ``edge'' of the 
Stream proper coincides with the increased number of \ion{H}{1} clouds
between $\ell\approx 98^\circ$ and $\ell\approx 104^\circ$.  

\placefigure{fig:image}

In the lower panel of 
Figure~6, we have highlighted three nearby ($\simlt$4$^\circ$
away from the line of sight) HVCs to the III~Zw~2 sightline.
While these HVCs are at
the threshold of the Leiden-Dwingeloo Survey detection
limit, each has been detected previously
(Hulsbosch \& Wakker 1988; Wakker \& van Woerden 1991):
HVC~104.0-51.0-337 was
detected by Hulsbosch \& Wakker (1988), but is subsumed into HVC\#493 by Wakker
\& van Woerden (1991); HVC~110.0-50.0-290 corresponds to HVC\#520 in Wakker
\& van Woerden; HVC~108.0-53.0-325 corresponds to HVC\#523.  Much of the
\ion{H}{1} detected with N(\ion{H}{1})$\simlt$10$^{18}$\,cm$^{-2}$ is
probably noise (and therefore the lower panel of
Figure~6 should be used
cautiously), but at least these three nearby HVCs have been
independently detected by Hulsbosch \& Wakker (1988).
More importantly, these neighboring HVCs demonstrate that gas
presumably associated with the Stream exists $\simgt 15^\circ$ 
away from the peak of the \ion{H}{1} emission.  Because the ``width'' of
the MS\,V concentration seen in the upper panel of Figure~6 is 
$\sim$10$^\circ$, the existence of MS
``froth'' beyond the outermost contour is not surprising.

Equipped only with our measurement of the \ion{Mg}{2} column density and the
upper limit
on the neutral hydrogen column density, it is difficult to constrain the
metallicity of the absorbing gas.  As an exploration of the possible
range in inferred
metallicity, we have constructed a grid of photoionization models
using the photoionization code CLOUDY (Version 90.04 - Ferland 1996).  We 
hope to constrain this range of
physical parameters more tightly with future 
observations.\footnote{Being radio-loud, III~Zw~2
may lend itself to potential future 21cm absorption analyses.}

We model the absorbing gas as a plane-parallel slab illuminated on one side
with a normally incident ionizing photon flux,
$\log\phi=5.5$\,photons\,cm$^{-2}$\,s$^{-1}$.  This is consistent with the
estimates of Bland-Hawthorn \& Maloney (1999a,b) 
for this region of the Magellanic
Stream, based on measurements of the H$\alpha$ emission near our line of
sight.  As Bland-Hawthorn \& Maloney discuss, this level of radiation is much
higher than that due to the 
general metagalactic background, $\log\phi= 4$ (Shull \etal\ 1999),
and is likely due to escaping
stellar radiation from the Milky Way.  As a result, we model the spectrum as
a power law with spectral index $\alpha_s=2$, assuming F$_\nu\propto\nu^{-2}$
between 1 and 4\,ryd, with a dropoff above 4\,ryd of a factor of 100.
This represents a rough approximation
to the integrated radiation from Galactic OB
associations, as well as a harder extragalactic
component.  However, adopting a T$_{eff}$ = 
35000\,K stellar atmosphere for the incident spectrum
yields a grid of models which are indistinguishable
from the following results.

For a given density n(H) and total hydrogen column density N(H),
the assumed metallicity of our cloud model was allowed to vary until
N(\ion{Mg}{2})=$(1.0\pm 0.05)\times 10^{13}$\,cm$^{-2}$ was
achieved.  The results of the grid of models are shown in Figure
7.  Also shown is the area that can be currently
excluded by the
requirement that N(\ion{H}{1})$<5\times 10^{18}$\,cm$^{-2}$
(shaded region).  
Our \ion{Mg}{2} column density determination, coupled with
the upper limit to N(\ion{H}{1}), allows us to set a lower limit
[\ion{Mg}{2}/\ion{H}{1}]$\simgt$$-$1.3.  Even this has only limited use
as a constraint on the metallicity [Mg/H], however, as for
N(\ion{H}{1})$<$$5\times 10^{18}$\,cm$^{-2}$ and
$\log\phi=5.5$, the ionization
correction for \ion{H}{1} is substantially more sensitive to changes
in density than that for \ion{Mg}{2}.

\placefigure{fig:cloudy_Mg}

\subsection{NGC~7469}
\label{analysis_n7469}

As already noted in \S~2.3, the Stream was detected in
\ion{Mg}{2}\,$\lambda$2796 in the FOS G270H spectrum of NGC~7469 (bottom panel
of Figure~8).  The
\ion{Mg}{2}\,$\lambda$2803 line lies below the 3$\sigma$ detection threshold and
is not discussed further.  The \ion{Mg}{2} column density inferred from 
the apparent optical depth method (Table~5)
is N(\ion{Mg}{2})$>(0.47\pm 0.11)\times 10^{13}$\,cm$^{-2}$.  
While the line does not \it appear \rm saturated, 
the inferior spectral resolution
of FOS ($\sim$230\,km\,s$^{-1}$), in comparison with GHRS
($\sim$19\,km\,s$^{-1}$), does not allow us to unequivocally rule out the
presence of saturation effects.
To be conservative, we assume the measured \ion{Mg}{2} column is a lower limit
and will revisit this sightline 
with our FUSE Science Team Cycle~1 observations of
NGC~7469, the \ion{O}{6} properties for which have already been discussed by
Sembach \etal\ (2000).

An NRAO
140$^\prime$ \ion{H}{1} spectrum (upper panel of Figure~8), 
kindly provided by Ken Sembach (to be
published in Murphy \etal\ 2000), 
shows that the \ion{H}{1} column density along this MS\,V sightline is
N(\ion{H}{1})=$(0.40\pm 0.04)\times 10^{19}$\,cm$^{-2}$.
These constraints lead to
\begin{equation}
[{\rm Mg~II/H~I}]>-1.51\pm 0.11({\rm r}).
\label{eq:n7469_Mg_final}
\end{equation}
\noindent
Errors in converting this measurement to an abundance
[Mg/H] are potentially large, since as discussed, for
N(\ion{H}{1}) $\approx 10^{18}$\,cm$^{-2}$, the ionization
correction for \ion{H}{1} is substantially more sensitive to density
than the correction for \ion{Mg}{2}.  For an
assumed density n(H)=0.1\,cm$^{-3}$, [\ion{Mg}{2}/\ion{H}{1}] may
overestimate [Mg/H] by an order of magnitude
or more.  In addition, the large uncertainty in
N(\ion{H}{1}) about this  $\sim$10$^{18}$\,cm$^{-2}$
range has the secondary effect of enhancing the
density sensitivity of the ionization correction to
\ion{H}{1}.  In general, however, ionization
corrections imply that [\ion{Mg}{2}/\ion{H}{1}] is an
overestimate of [Mg/H], ranging from a factor of 2
to 10 for n(H) ranging from (1$-$0.1)\,cm$^{-3}$.

\placefigure{fig:n7469_stack}

\placetable{tbl:n7469_lines}

\section{Discussion}
\label{discussion}

\subsection{Fairall~9}
\label{discussion_f9}

Based on its proximity to the SMC and the
disrupted appearance of the SMC, it is likely that the MS\,I (and
the Stream as a whole) originated in the SMC.  
If the Stream was drawn out of the SMC
$\sim$1.5\,Gyrs ago, as the best tidal models predict (Gardiner 1999), one
would expect the metallicity to reflect that of the SMC at that epoch. 
According to the best chemical evolution models (Pagel \& Tautvaisiene 1998;
Figure~5), that
should actually be $\sim$0.2\,dex smaller
than the present-day value of
[S/H]=$-$0.68 (Russell \& Dopita 1992).
Note that this argument
also holds for an LMC origin as well; the LMC metallicity 1.5\,Gyrs ago, 
in the mean, is also predicted to be $\sim$0.2\,dex 
lower than the present-day LMC
value of [S/H]=$-$0.57
(Pagel \& Tautvaisiene; Figure 4).  In either scenario, the
predicted [S/H] would appear to be mildly inconsistent with our observations,
[\ion{S}{2}/\ion{H}{1}]=$-$0.55$\pm$0.06(r)$\pm$$\,\,^{+0.17}_{-0.21}$(s).
On the other hand,
the stochastic nature of star formation and the magnitude of the
observational scatter (see Figures 4 and 5 of Pagel \& Tautvaisiene),
severely limits
the predictive power of even the best Magellanic Cloud chemical evolution
models, applied to the most recent $1-2$\,Gyrs. 
Thus, this apparent discrepancy should not be overinterpreted.

An inherent assumption in the above picture is that the MS\,I gas can be
linked directly to disk gas-phase abundances.  The present-day
population of disk \ion{H}{2} regions, in both the LMC and SMC, shows no
evidence for any substantial abundance gradients (Dufour 1975), nor are the
gaseous disks of the Clouds (currently) substantially larger than their
optical disks.  In other words, based on the present state of the
Magellanic Clouds, there is no reason to suspect that the Stream abundances should
not reflect the disk gas-phase abundances.  A caveat to this statement
is that the present-day conditions may not necessarily reflect those
when the Stream
was purported to form $\sim$1.5\,Gyrs ago.  
Even if an abundance gradient or extended gaseous disk existed when the
Stream formed, unless the gradient was inverted (highly unusual), one would
then expect the Stream abundances to be even lower than the
disk gas-phase abundances.  This would worsen the comparison of equation
6 with either the present-day Russell \& Dopita (1992) LMC/SMC
abundances, or those corrected downward by 0.2\,dex based upon conservative
chemical evolution model predictions.  
Higher spatial resolution \ion{H}{1}
data will be required to reduce the remaining substantial systematic
uncertainties discussed in \S~3.1, and to better ascertain whether
a legitimate discrepancy exists between the MS\,I metallicity and that expected
for an SMC disk origin. 
Impending synthesis data from our Australia Telescope
Compact Array program (Putman \& Gibson 2000) should aid in this 
regard.

Complicating the above interpretation is the recent work of
Rolleston \etal\ (1999), who found
that, for three B-stars in the Magellanic Bridge, [Mg/H]=$-0.94\pm 0.14$ and
[Si/H]=$-1.23\pm 0.25$.  These Bridge stars appear to be $\sim$0.5\,dex
deficient in metals compared to a similar
B-star (AV~304) in the SMC itself.  These stars
do not appear to have formed from gas of the present-day SMC or LMC
composition.
Some unenriched component most likely mixed with the SMC gas to
yield abundances low enough to 
produce the Rolleston \etal\ results.  Evidence is
presented therein for the SMC not being as well-mixed as Dufour (1975) 
had claimed earlier, potentially providing a source for metal-poor
contaminating gas.  Invoking this metal-poor
contamination is counter to that implied by our Fairall~9 MS\,I sightline,
which appears to be enriched
(by 0.1$-$0.2\,dex), in comparison with the present-day mean
gas-phase abundance of the SMC.

Equations 6 and 7 imply
[Si/S]$\simgt$$-$1.
While this lower limit marginally excludes dust depletion of the
magnitude seen in cool diffuse Galactic disk clouds (Savage \& Sembach 1996;
Figure~6), it is not sufficiently
restrictive to discriminate between depletion of the magnitude seen in
warm Galactic halo clouds (as might be expected for an SMC
origin for the Stream gas -- Welty \etal\ 1997) and
warm Galactic disk clouds (as might be expected for an LMC origin -- Welty 
\etal\ 1999), which typically show silicon depletion of order
[Si/S]$\approx$$-$0.4$\pm$0.1 (Savage \& Sembach 1996; Figure~6).
Further, it is known that the gas-phase in 
both the LMC and SMC is relatively overabundant in
silicon, showing [Si/S]$\approx$$+$0.08$\pm$0.04 (Welty \etal\ 1997,1999).
In other words, our limit limit of [Si/S]$\simgt$$-$1 for the Magellanic Stream
cannot be used as a constraint to differentiate between an LMC versus SMC
origin.  Impending FUSE Cycle~1 Science Team observations of Fairall~9
(scheduled for July 2000), when coupled with our GHRS data,
should shed further light on the dust depletion pattern of the Magellanic
Stream.

\subsection{III~Zw~2}
\label{discussion_mrk1501}

Assuming that the gas near MS\,V, $\sim$80$^\circ$ down-Stream from
Fairall~9, originates from within the Magellanic
Clouds, it is tempting to assign it a magnesium abundance
that reflects that of either the LMC or SMC.  Unfortunately, the gas-phase
magnesium abundance for both Clouds is highly uncertain.  Only for the LMC is
there a direct measure of the interstellar medium (ISM) gas-phase magnesium,
through the IUE absorption line analysis of R136 (de~Boer \etal\ 1985;
Welty \etal\ 1999), the result for which was [Mg/H]$=$$-$1.1$\pm$0.3.  In
contrast, spectral synthesis of B-star and F-type supergiant atmospheres
implies the much higher abundance [Mg/H]$\approx$$-$0.5$\rightarrow$$-$0.1
(Welty \etal\ 1999; Table~11).  For the SMC, no direct ISM absorption line
abundances have been published yet, but B-star and supergiant spectral
synthesis studies lead to [Mg/H]$\approx$$-$0.7$\rightarrow$$-$0.4
(Welty \etal\ 1997; Rolleston \etal\ 1999).

As the model curves of Figure
7 show, neither an LMC nor an SMC origin can
be excluded based upon our knowledge of N(\ion{Mg}{2}) and N(\ion{H}{1}),
as even the most extreme allowable values for the LMC and SMC magnesium
abundances (see above) reside in the parameter space not yet excluded by the
upper limit to N(\ion{H}{1}) -- \ie, the shaded region.
In either case, the total hydrogen column along this sightline would be
dominated by N(\ion{H}{2}) and not N(\ion{H}{1}), and the total magnesium
column would be dominated by N(\ion{Mg}{3}).  This latter result should
not be surprising, since the ionization potentials of \ion{H}{1} and
\ion{Mg}{2} lie within 10\% of each other.  Where hydrogen is
predominantly ionized, magnesium follows suit, although as discussed
above, more magnesium remains in the singly ionized state.

From the Leiden-Dwingeloo Survey (Hartmann \& Burton 1997), we can derive the
\ion{H}{1} volume density for both the
marginally resolved HVC\,100.0-48.5-390 (Figure~6) and the
clearly resolved MS\,V concentration at $(\ell,b)\approx
(92^\circ,-51^\circ)$.  In both cases, \ion{H}{1} column densities
$\sim 2\times 10^{19}$\,cm$^{-2}$ are found.  If the radial
extent of the clouds is assumed to be similar to their lateral extent,
the angular width times the distance to the clouds must be
of order N(\ion{H}{1})/n(\ion{H}{1}).
At the assumed distance of the LMC
($\sim$50\,kpc), this corresponds to an \ion{H}{1} volume
density of n(\ion{H}{1})$\approx 10^{-1.7}$cm$^{-3}$.  
If one were to make the
assumption that this \ion{H}{1} volume density truly
reflected the total hydrogen volume density in
those higher column density clouds, and that this
represented the typical total hydrogen density of this part of the
Magellanic Stream,
this would imply that the total hydrogen column density 
for the absorber toward III Zw 2 was
N(H)=$(1-4)\times 10^{18}$\,cm$^{-2}$ 
with N(\ion{H}{2})$\gg $N(\ion{H}{1}).

Because we have no definitive evidence of the nature (if any) of the dust
depletion pattern in the Magellanic Stream, the above CLOUDY analysis
necessarily adopted the simplifying assumption
of zero dust -- \ie, pure photoionization
effects were employed, in an attempt to reconcile the observed \ion{Mg}{2} and
\ion{H}{1} Stream constraints with the observed [Mg/H] of the LMC and SMC.
Such an analysis is a useful exercise, but is no doubt an over-simplification,
due to the potential complicating effects of dust depletion.
If we subscribe to the
conclusion of \S~4.1, that the Stream dust depletion
may range anywhere from nonexistent, 
to that seen in warm diffuse Galactic clouds,
then our measured gas-phase \ion{Mg}{2} abundance may
underestimate the true magnesium abundance by up to a factor of ten.  
More sensitive \ion{H}{1} observations, coupled
with additional absorption measurements for other ions, are necessary.

The LMC and SMC possess magnesium abundances of 
[Mg/H]$\approx$$-$1.1$\rightarrow$$-$0.1 and
[Mg/H]$\approx$$-$0.7$\rightarrow$$-$0.4, respectively.  Regardless of the
origin of the high-velocity Stream, it is likely that
some degree of dust depletion will be present, since both the LMC and SMC
exhibit distinct depletion patterns; the former displays a pattern similar to
that seen in warm Galactic
disk clouds (Welty \etal\ 1999), while the pattern seen in the 
latter resembles that of warm Galactic halo clouds (Welty \etal\ 1997).
Our silicon analysis of the
Fairall~9 sightline (\S~3.1) showed that we could not
exclude dust depletion
patterns similar to those of warm diffuse Galactic gas clouds.  This implies
a potential correction of $\sim$0$\rightarrow$$+$1\,dex
to the measured magnesium abundance (Savage \& Sembach 1996; Figure~6).
This would retain the consistency of the inferred
[\ion{Mg}{2}/\ion{H}{1}] limits of \it both \rm III~Zw~2 
([\ion{Mg}{2}/\ion{H}{1}]$\simgt$$-$1.3) and NGC~7469 
([\ion{Mg}{2}/\ion{H}{1}]$\simgt$$-$1.5) with
the present-day SMC stellar abundance
value of [Mg/H]$\approx$$-$0.7$\rightarrow$$-$0.4.

\section{Summary}
\label{summary}

We summarize in Table~6 the
results of our column density measurements and limits for
three sightlines through the Magellanic Stream.
We have also included our inferences
about the abundances of the gas seen along these sightlines,
many of which remain uncertain.  For the
line of sight toward Fairall~9, where ionization corrections
are expected to be small, measurements of [\ion{S}{2}/\ion{H}{1}] are
consistent with the present-day abundances of both the LMC and SMC.
If this gas was tidally-stripped from the SMC 1.5\,Gyrs ago, its
abundance is approximately a factor of two greater than that expected.
Stochastic star formation
effects and remaining systematic uncertainties (particularly in the \ion{H}{1}
column density) could weaken this statement.
New HST/STIS and FUSE observations of Fairall~9 at \ion{S}{3} will test
the inherent assumption that \ion{S}{3} ionization corrections are
negligible.  Similarly, both III~Zw~2 and NGC~7469 
need to be revisited with HST/STIS, in
order to derive \ion{S}{2} and \ion{S}{3} column densities.  NGC~7469
has been observed by FUSE and its \ion{O}{6} properties discussed by
Sembach \etal\ (2000); further analysis of this FUSE dataset is 
currently underway.  Fairall~9 will likewise be observed by the FUSE Science
Team in July 2000.

\placetable{tbl:results}

A fourth potential probe of the Magellanic Stream, in a high \ion{H}{1} column
density region of MS\,III, is NGC~7714.  It was observed with GHRS by 
Gonz\'alez-Delgado \etal\ (1999), but with
the low-resolution G140L grating.  The resulting resolution
($>$100\,km\,s$^{-1}$) was insufficient to accurately separate MS\,III
absorption features from the saturated Galactic lines, as the expected
separation (based upon the LDS \ion{H}{1} spectrum for this sightline) is only
$\sim$50\,km\,s$^{-1}$.  Scheduled FUSE 
observations of NGC~7714 should resolve
any far-UV MS\,III lines.

\acknowledgments

This work was supported by the NASA HST General Observer Grants GO-06586.01-95A
and GO-06593.01-95A, NASA Long-Term Space
Astrophysics Program (NAG5-7262), NSF grant
AST 96-17073,  
and the FUSE Science Team (NAS5-32985).
It was based in part on observations with the NASA/ESA
{\it Hubble Space Telescope} obtained at the Space Telescope Science
Institute, which is operated by AURA, Inc., under NASA contract
NAS5-26555.  We wish to
thank Phil Maloney and Bart Wakker for helpful discussions, Ken Sembach 
and Lister Staveley-Smith for providing
the NGC~7469 and Fairall~9, respectively, neutral hydrogen data
prior to publication, and
Gary Ferland for the use of CLOUDY.  A special thanks is due the referee (Chris
Howk), whose detailed report improved greatly the quality of the
final manuscript.


\clearpage

\noindent
Fig. 1. -- 
HST GHRS spectrum of Fairall~9 taken with the 
G160M grating.  Fairall~9 samples Magellanic Stream
gas associated with the MS\,I concentration 
(see Figure 1 of Mathewson 1985).  The raw spectrum has been smoothed by the
inverse of the 
post-COSTAR, GHRS, large science aperture, line spread function (LSF --
Gilliland
1994), in order to improve the S/N (at the admitted expense of some resolution)
and eliminate any LSF-induced asymmetries in the line profiles
(Penton, Stocke \& Shull 2000).
Absorption lines from both Galactic (G) and Magellanic Stream (MS)
S\,II (at 1250 and 1253\,\AA) and Si\,II (at 1260\,\AA)
are seen and labeled accordingly.  Galactic 
S\,II\,$\lambda$1259
is also seen, but since the corresponding Stream line is blended with Galactic
Si\,II\,$\lambda$1260,
we exclude it from our analysis.  Similarly, we
will not discuss further the $z=0.032$ Ly$\alpha$ line seen at
$\lambda=1254.15$\,\AA.

\noindent
Fig. 2. -- 
HST GHRS spectrum of III~Zw~2 taken with the G270M
grating.  III~Zw~2 lies $\sim$15$^\circ$ from the peak of the \ion{H}{1}
emission associated with the MS\,V concentration of the Magellanic Stream
(see Figure 1 of Mathewson 1985).  The raw spectrum has been smoothed by the
inverse of the
post-COSTAR LSF (Gilliland 1994), as noted in the caption to
Figure~1.
Both Galactic (G) and Magellanic Stream (MS) 
\ion{Mg}{2}~$\lambda\lambda$~2796,2803\AA\ 
is seen in absorption.  The centroid
of the Galactic
lines are blueshifted 0.24\,\AA\ with respect to their rest wavelengths
(Pickering, Thorne \& Webb 1998), reflecting the blended two-component 
structure of the Galactic gas along this sightline.

\noindent
Fig. 3. -- 
HST FOS spectrum of NGC~7469 taken with the G270H
grating.  
Galactic (G) and Magellanic Stream (MS) 
\ion{Mg}{2}~$\lambda\lambda$~2796,2803\AA\ 
is seen in absorption, although the Stream is clearly
detected in Mg\,II\,$\lambda$2796 only.

\noindent
Fig. 4. -- 
Velocity stack showing \ion{H}{1} (upper panel),
\ion{S}{2} (middle panels) and \ion{Si}{2} (lower panel) Galactic (G) and
Magellanic Stream (MS) features along the Fairall~9 sightline.  The
Hanning-smoothed \ion{H}{1} spectrum was collected with
the Parkes Multibeam Narrowband System (Haynes \etal\ 1999) and
was transformed from the heliocentric to
the local standard of rest frame
via $v_{\rm LSR}=v_\odot-11.6$\,km\,s$^{-1}$.  A
mild fourth-order polynomial baseline was subtracted from the raw spectrum,
and the native Jy/beam converted to T$_{\rm B}$ via
a multiplicative scale factor of 0.82 (Staveley-Smith 1999).  The
corresponding \ion{H}{1} column densities for both the Galactic and
Magellanic Stream components are labeled.

\noindent
Fig. 5. -- 
Velocity stack showing \ion{H}{1} (upper panel)
and \ion{Mg}{2} (middle and lower panels) Galactic (G) and
Magellanic Stream (MS) features along the III~Zw~2 sightline.  The
Hanning-smoothed \ion{H}{1} spectrum was taken from the Leiden-Dwingeloo Survey
(Hartmann \& Burton 1997).

\noindent
Fig. 6. -- 
\it Upper Panel: \rm 
Contours of neutral hydrogen column density 
N(\ion{H}{1}) in the vicinity of the MS\,V component of the Magellanic Stream.
Contour levels are $5\times 10^{18}$\,cm$^{-2}$, with the outermost level
N(\ion{H}{1})=$5\times 10^{18}$\,cm$^{-2}$ over the velocity range 
$-400<v_{\rm LSR}<-250$\,km\,s$^{-1}$.  The \ion{H}{1}
data were taken from the
Leiden-Dwingeloo Survey (Hartmann \& Burton 1997); the spatial resolution is
restricted by the 1/2-degree sampling grid (with a 1/2-degree beam) employed.
Both the III~Zw~2 and
NGC~7469 sightlines are noted.  The peak of the H\,I emission
associated with MS\,V lies near ($\ell$,$b$)$\approx$($92^\circ$,$-51^\circ$).
\it Lower Panel: \rm Neutral hydrogen in the immediate vicinity (\ie, the
area covered by the marked box in the upper panel) of our III~Zw~2 sightline.  
Contour levels are now $3\times 10^{18}$\,cm$^{-2}$, with the outermost level 
N(\ion{H}{1})$=1\times 10^{18}$\,cm$^{-2}$, over the velocity range 
$-400<v_{\rm LSR}<-250$\,km\,s$^{-1}$.
The local standard of rest
velocities of three independently-confirmed (Hulsbosch \& Wakker 1988)
High-Velocity Clouds within 4$^\circ$ of our III~Zw~2 sightline are
labeled.  The lack of detectable H\,I at the position and velocity
of our Mg\,II detection in the spectrum of III~Zw~2 sets an upper limit
of N(\ion{H}{1})=$5\times 10^{18}$\,cm$^{-2}$.

\noindent
Fig. 7. -- 
Grid of CLOUDY 
(Version 90.04 - Ferland 1996) models
showing contours of total magnesium to total hydrogen ([Mg/H]) as a function of
total hydrogen column N(H) and volume n(H) densities.  The 
models satisfy the observational constraints that N(\ion{Mg}{2})$\equiv 1\times
10^{13}$\,cm$^{-2}$ and N(\ion{H}{1})$<5\times 10^{18}$\,cm$^{-2}$,
representative constraints imposed by our III~Zw~2 analysis.  
A normally-incident ionizing
photon flux of $\phi=3\times 10^5$\,photons\,cm$^{-2}$\,s$^{-1}$ was
assumed, consistent with that expected at MS\,V, according to the models of
Bland-Hawthorn \& Maloney (1999a,b).  
The shaded region corresponds to the (unfortunately) small part of
parameter space excluded by the \ion{H}{1} column density constraint.

\noindent
Fig. 8. -- 
Velocity stack showing \ion{H}{1} (upper panel)
and \ion{Mg}{2}\,$\lambda$2796 (middle and lower panels) Galactic (G) and
Magellanic Stream (MS) features along the III~Zw~2 sightline.  The
Hanning-smoothed \ion{H}{1} spectrum was collected at the NRAO 140$^\prime$
and kindly provided by Ken Sembach prior to publication (Murphy, Sembach
\& Lockman 2000).

\clearpage

\begin{deluxetable}{llclrccccc}
\footnotesize
\tablecaption{HST Observations \label{tbl:obs}}
\tablewidth{0pt}
\tablehead{
\colhead{PID/PI\tablenotemark{a}} &
\colhead{Dataset\tablenotemark{b}} &
\colhead{Date} &
\colhead{Target} &
\colhead{t$_{\rm exp}$\tablenotemark{c} [s]} &
\colhead{Instrument} &
\colhead{Grating} &
\colhead{Resolution} &
\colhead{$\lambda_i$} &
\colhead{$\lambda_f$} \nl
\colhead{} &
\colhead{} &
\colhead{} &
\colhead{} &
\colhead{} &
\colhead{} &
\colhead{} &
\colhead{[km\,s$^{-1}$]} &
\colhead{[\AA]} &
\colhead{[\AA]}
}
\startdata
5300/Savage & z26o0208n & 04/09/94 & Fairall~9 &  7096.3 & GHRS & G160M & 
  19 & 1231.7 & 1267.8 \nl
6593/Stocke & z3e70404m & 08/01/96 & Fairall~9 & 13407.0 & GHRS & G160M & 
  19 & 1219.7 & 1255.8 \nl
6593/Stocke & z3e70406t & 08/01/96 & Fairall~9 &  6412.0 & GHRS & G160M & 
  19 & 1240.3 & 1276.4 \nl
6586/Stocke & z3j45104t & 01/03/97 & III~Zw~2  &  1501.0 & GHRS & G270M & 
  19 & 2786.2 & 2833.3 \nl
6747/Kriss  & y3b6010bt & 06/18/96 & NGC~7469  &   300.0 & FOS  & G270H & 
 230$\;\;$ & 2221.6 & 3300.7 \nl
\enddata
\tablenotetext{a}{\vspace{-6.0mm}HST Proposal ID and Principal Investigator.}
\tablenotetext{b}{\vspace{1.0mm}HST Archive dataset filename.}
\tablenotetext{c}{The signal-to-noise ratio per resolution element
for the merged Fairall~9 spectrum employed in our analysis is S/N=30 (at
$\lambda=1250$\,\AA).  For the single III~Zw~2 spectrum, S/N=8 (at
$\lambda=2800$\,\AA); for the single NGC~7469 spectrum, S/N=29 (at
$\lambda=2800$\,\AA).
The 3$\sigma$ minimum equivalent width
detectable for the merged Fairall~9 spectrum
is 6\,m\AA\ (at $\lambda=1250$\,\AA).  For the III~Zw~2
spectrum, 27\,m\AA\ (at $\lambda=2800$\,\AA), and for the NGC~7469 spectrum,
160\,m\AA\ (at $\lambda=2800$\,\AA).}
\end{deluxetable}

\clearpage

\begin{deluxetable}{ccrccc}
\footnotesize
\tablecaption{Galactic and Magellanic Stream Absorption Features in Fairall~9
Spectrum
\label{tbl:F9_lines}}
\tablewidth{0pt}
\tablehead{
\colhead{$\lambda_c$\tablenotemark{a}} &
\colhead{$\Delta v_{{\rm LSR}}$\tablenotemark{b}} &
\colhead{W$_\lambda$} &
\colhead{N$_{\tau=0}$\tablenotemark{c}} &
\colhead{N$_{\tau_v}$\tablenotemark{d}} &
\colhead{ID} \nl
\colhead{[\AA]} &
\colhead{[km\,s$^{-1}$]} &
\colhead{[m\AA]} &
\colhead{[cm$^{-2}$]} &
\colhead{[cm$^{-2}$]} &
\colhead{}
}
\startdata
1250.6  &   $-$55$\rightarrow$$+$45   &   75$\pm$4   &
	            (1.04$\pm$0.06)$\times$10$^{15}$ &
	            (1.25$\pm$0.06)$\times$10$^{15}$ &
	   \ion{S}{2}\,$\lambda$1250: Galaxy \nl
1251.4  &  $+$125$\rightarrow$$+$250  &   36$\pm$4   &
	            (5.00$\pm$0.56)$\times$10$^{14}$ &
	            (5.39$\pm$0.61)$\times$10$^{14}$ &
	   \ion{S}{2}\,$\lambda$1250: Stream \nl
1253.8  &   $-$55$\rightarrow$$+$45   &  114$\pm$3   &
	            (0.80$\pm$0.02)$\times$10$^{15}$ &
	            (1.03$\pm$0.03)$\times$10$^{15}$ &
	   \ion{S}{2}\,$\lambda$1253: Galaxy \nl
1254.6  &  $+$125$\rightarrow$$+$250  &   62$\pm$5   &
	            (4.33$\pm$0.35)$\times$10$^{14}$ &
	            (4.75$\pm$0.41)$\times$10$^{14}$ &
	   \ion{S}{2}\,$\lambda$1253: Stream \nl
1260.5\tablenotemark{e}  &  $-$150$\rightarrow$$+$85   &  427$\pm$5   &
	            (2.57$\pm$0.03)$\times$10$^{13}$ &
 	                    $\qquad\;\;>$5.44$\times$10$^{13}$ &
	   \ion{Si}{2}\,$\lambda$1260: Galaxy \nl
1261.2  &   $+$85$\rightarrow$$+$350  &  561$\pm$5   &
	            (3.38$\pm$0.03)$\times$10$^{13}$ &
 	                    $\qquad\;\;>$9.29$\times$10$^{13}$ &
	   \ion{Si}{2}\,$\lambda$1260: Stream \nl
\enddata
\tablenotetext{a}{Line centroid.}
\tablenotetext{b}{Velocity range over which the spectral
line integration was applied.}
\tablenotetext{c}{Inferred column density, under the assumption that
the line in question is optically thin.}
\tablenotetext{d}{Inferred column density, employing the $\tau_v$ technique
of Sembach \& Savage (1992), neglecting continuum placement
uncertainties.}
\tablenotetext{e}{Galactic
\ion{Fe}{2}\,$\lambda$1260 and \ion{C}{1}\,$\lambda$1260 appear at
$+$26\,km\,s$^{-1}$ and $+$75\,km\,s$^{-1}$, respectively, in this reference
frame.  This will impact the lower limit on the column density for the Galactic
\ion{Si}{2}\,$\lambda$1260 line, and to a lesser extent, the Stream
\ion{Si}{2}\,$\lambda$1260 line.}
\end{deluxetable}

\clearpage
\begin{deluxetable}{llll}
\footnotesize
\tablecaption{Atomic Data 
\label{tbl:atomic}}
\tablewidth{0pt}
\tablehead{
\colhead{ID} &
\colhead{$\lambda_0$\tablenotemark{a} [\AA]} &
\colhead{$f$\tablenotemark{b}} &
\colhead{$\qquad\qquad$A$_{{\rm X}_{\odot}}$\tablenotemark{c}}
}
\startdata
\ion{S}{2}\,$\lambda$1250  & 1250.578  & 0.00520 & $(1.862\pm 0.215)\times 10^{-5}$ \nl
\ion{S}{2}\,$\lambda$1253  & 1253.805  & 0.0103  & $(1.862\pm 0.215)\times 10^{-5}$\nl
\ion{Si}{2}\,$\lambda$1260 & 1260.4221 & 1.18    & $(3.548\pm 0.164)\times 10^{-5}$ \nl
\ion{Mg}{2}\,$\lambda$2796 & 2796.354  & 0.629   & $(3.802\pm 0.175)\times 10^{-5}$ \nl
\ion{Mg}{2}\,$\lambda$2803 & 2803.531  & 0.314   & $(3.802\pm 0.175)\times 10^{-5}$\nl
\enddata
\tablenotetext{a}{Rest wavelengths from Verner, Barthel \& Tytler (1994),
except \ion{Mg}{2}, which are from Pickering \etal\ (1998).}
\tablenotetext{b}{Oscillator strength from Verner \etal\ (1994).}
\tablenotetext{c}{Solar (meteoritic) abundance from Table 2 of
Anders \& Grevesse (1989), for the element X (where X corresponds to S, Si, and
Mg, here).}
\end{deluxetable}

\clearpage

\begin{deluxetable}{cccccc}
\footnotesize
\tablecaption{Galactic and Magellanic Stream Absorption Features in III~Zw~2
Spectrum
\label{tbl:IIIZw2_lines}}
\tablewidth{0pt}
\tablehead{
\colhead{$\lambda_c$\tablenotemark{a}} &
\colhead{$\Delta v_{{\rm LSR}}$\tablenotemark{b}} &
\colhead{W$_\lambda$} &
\colhead{N$_{\tau=0}$\tablenotemark{c}} &
\colhead{N$_{\tau_v}$\tablenotemark{d}} &
\colhead{ID} \nl
\colhead{[\AA]} &
\colhead{[km\,s$^{-1}$]} &
\colhead{[m\AA]} &
\colhead{[cm$^{-2}$]} &
\colhead{[cm$^{-2}$]} &
\colhead{}
}
\startdata
2793.1  &  $-$400$\rightarrow$$-$320  &  315$\pm$45  &  
	   (0.72$\pm$0.10)$\times$10$^{13}$ &
	   (1.08$\pm$0.10)$\times$10$^{13}$ &
	   \ion{Mg}{2}\,$\lambda$2796: Stream \nl
2796.2  &  $-$100$\rightarrow$$+$50   &  765$\pm$47  &  
	   (1.76$\pm$0.11)$\times$10$^{13}$ &
 	 $\qquad\;\;>$3.39$\times$10$^{13}$ &
	   \ion{Mg}{2}\,$\lambda$2796: Galaxy \nl
2800.3  &  $-$400$\rightarrow$$-$320  &  149$\pm$63  & 
           (0.68$\pm$0.29)$\times$10$^{13}$ &
           (0.91$\pm$0.14)$\times$10$^{13}$ &
	   \ion{Mg}{2}\,$\lambda$2803: Stream \nl
2803.3  &  $-$100$\rightarrow$$+$50   &  816$\pm$78  & 
           (3.74$\pm$0.36)$\times$10$^{13}$ &
 	 $\qquad\;\;>$7.54$\times$10$^{13}$ &
	   \ion{Mg}{2}\,$\lambda$2803: Galaxy \nl
\enddata
\tablenotetext{a}{Line centroid.}
\tablenotetext{b}{Velocity range over which the spectral
line integration was applied.}
\tablenotetext{c}{Inferred column density, under the assumption that
the line in question is optically thin.}
\tablenotetext{d}{Inferred column density, employing the $\tau_v$ technique
of Sembach \& Savage (1992), neglecting continuum placement
uncertainties.}
\end{deluxetable}

\clearpage

\begin{deluxetable}{ccccrc}
\footnotesize
\tablecaption{Galactic and Magellanic Stream Absorption Features in NGC~7469
Spectrum
\label{tbl:n7469_lines}}
\tablewidth{0pt}
\tablehead{
\colhead{$\lambda_c$\tablenotemark{a}} &
\colhead{$\Delta v_{{\rm LSR}}$\tablenotemark{b}} &
\colhead{W$_\lambda$} &
\colhead{N$_{\tau=0}$\tablenotemark{c}} &
\colhead{N$_{\tau_v}$\tablenotemark{d}} &
\colhead{ID} \nl
\colhead{[\AA]} &
\colhead{[km\,s$^{-1}$]} &
\colhead{[m\AA]} &
\colhead{[cm$^{-2}$]} &
\colhead{[cm$^{-2}$]} &
\colhead{}
}
\startdata
2793.4  &  $-$500$\rightarrow$$-$200  &  191$\pm$42  &  
	   (0.44$\pm$0.10)$\times$10$^{13}$ &
           $>$(0.47$\pm$0.11)$\times$10$^{13}$ &
	   \ion{Mg}{2}\,$\lambda$2796: Stream \nl
2796.9  &  $-$200$\rightarrow$$+$300  &  652$\pm$57  & 
           (1.50$\pm$0.13)$\times$10$^{13}$ &
           (1.67$\pm$0.16)$\times$10$^{13}$ &
	   \ion{Mg}{2}\,$\lambda$2796: Galaxy \nl
2803.7  &  $-$150$\rightarrow$$+$250  &  520$\pm$51  & 
           (2.38$\pm$0.23)$\times$10$^{13}$ &
           (2.65$\pm$0.27)$\times$10$^{13}$ &
	   \ion{Mg}{2}\,$\lambda$2803: Galaxy \nl
\enddata
\tablenotetext{a}{Line centroid.}
\tablenotetext{b}{Velocity range over which the spectral
line integration was applied.}
\tablenotetext{c}{Inferred column density, under the assumption that
the line in question is optically thin.}
\tablenotetext{d}{Inferred column density, employing the $\tau_v$ technique
of Sembach \& Savage (1992), neglecting continuum placement
uncertainties.}
\end{deluxetable}

\clearpage

\begin{deluxetable}{lcccrcrr}
\footnotesize
\tablecaption{Summary of Magellanic Stream Abundance 
Determinations\tablenotemark{a}
\label{tbl:results}}
\tablewidth{0pt}
\tablehead{
\colhead{Probe} &
\colhead{N(\ion{S}{2})\tablenotemark{b}} &
\colhead{N(\ion{Si}{2})\tablenotemark{c}} &
\colhead{N(\ion{Mg}{2})\tablenotemark{d}} &
\colhead{N(\ion{H}{1})} &
\colhead{[\ion{S}{2}/\ion{H}{1}]} &
\colhead{$\qquad$[\ion{Si}{2}/\ion{H}{1}]} &
\colhead{[\ion{Mg}{2}/\ion{H}{1}]} \nl
\colhead{} &
\colhead{[$10^{14}$\,cm$^{-2}$]} &
\colhead{[$10^{13}$\,cm$^{-2}$]} &
\colhead{[$10^{13}$\,cm$^{-2}$]} &
\colhead{[$10^{19}$\,cm$^{-2}$]} &
\colhead{} &
\colhead{} &
\colhead{} \nl
}
\startdata
Fairall~9    & $4.95\pm 0.34$                      & 
	    $>$9.29                                &
			                           & 
	       $9.35\pm 0.47$                      &  
	      $-0.55\pm 0.06$                      & 
	   $>$$-$1.55$\pm$0.03                     & 
			                           \nl
III~Zw~2  &                                        &
						   & 
	    $1.02\pm 0.08$                         & 
	    $<0.5\qquad\quad\;\;$                  &
						   &
						   & 
        $>$$-$1.27$\pm$0.04                        \nl  
NGC~7469  &                                        &
						   & 
	    $>$0.47$\pm$0.11$\;\;\;\;\,$              & 
	    $0.40\pm 0.04$                         &
						   &
						   & 
	   $>$$-$1.51$\pm$0.11                     \nl  
LMC\tablenotemark{e}&&&&& $-0.57\pm 0.10$ & $-$0.53$\pm$0.22 & 
	\sl $-$1.1$\rightarrow$$-$0.1\rm \nl
SMC\tablenotemark{e}&&&&& $-0.68\pm 0.16$ & $-$0.57$\pm$0.17 &  
	\sl $-$0.7$\rightarrow$$-$0.4\rm \nl
\enddata
\tablenotetext{a}{Column densities N(\ion{S}{2}), N(\ion{Si}{2}), and 
N(\ion{Mg}{2}) reflect those derived using the $\tau_v$ technique
(Sembach \& Savage 1992).  Quoted uncertainties correspond
to the total statistical
noise (equation A27 of Sembach \& Savage 1992); continuum placement 
uncertainties are not included here.}
\tablenotetext{b}{Based on the weighted mean of N(\ion{S}{2}\,$\lambda$1250) 
and N(\ion{S}{2}\,$\lambda$1253).}
\tablenotetext{c}{Based on the saturated line of \ion{Si}{2}\,$\lambda$1260.}
\tablenotetext{d}{Based on the weighted mean of 
N(\ion{Mg}{2}\,$\lambda$2796) and
N(\ion{Mg}{2}\,$\lambda$2803).  In the case of NGC~7469, only
N(\ion{Mg}{2}\,$\lambda$2796) was considered.}
\tablenotetext{e}{Gas-phase abundances -- \ion{S}{2} from Russell \& 
Dopita (1992); \ion{Si}{2} for the LMC from Welty \etal\
(1999); \ion{Si}{2} for the SMC from Welty \etal\ (1997).
The gas-phase magnesium abundances for both the LMC and SMC
are \it highly \rm uncertain.  For the LMC, IUE interstellar medium (ISM)
absorption line analyses along the line of sight to
R136 imply [Mg/H]=$-$1.1$\pm$0.3 (Welty \etal\ 1999; de Boer \etal\ 1985);
analyses of supernova remnants, B-stars, and supergiants
all yield [Mg/H]$\approx$$-$0.5$\rightarrow$$-$0.1 
(Welty \etal\ 1999; Table~11).
For the SMC, no ISM absorption line analyses have been undertaken;
spectral synthesis of B-star and supergiant atmospheres yield 
[Mg/H]$\approx$$-$0.7$\rightarrow$$-$0.4 (Rolleston \etal\ 1999;
Welty \etal\ 1997; Table~9).}
\end{deluxetable}

\clearpage

\epsscale{1.0}
\plotone{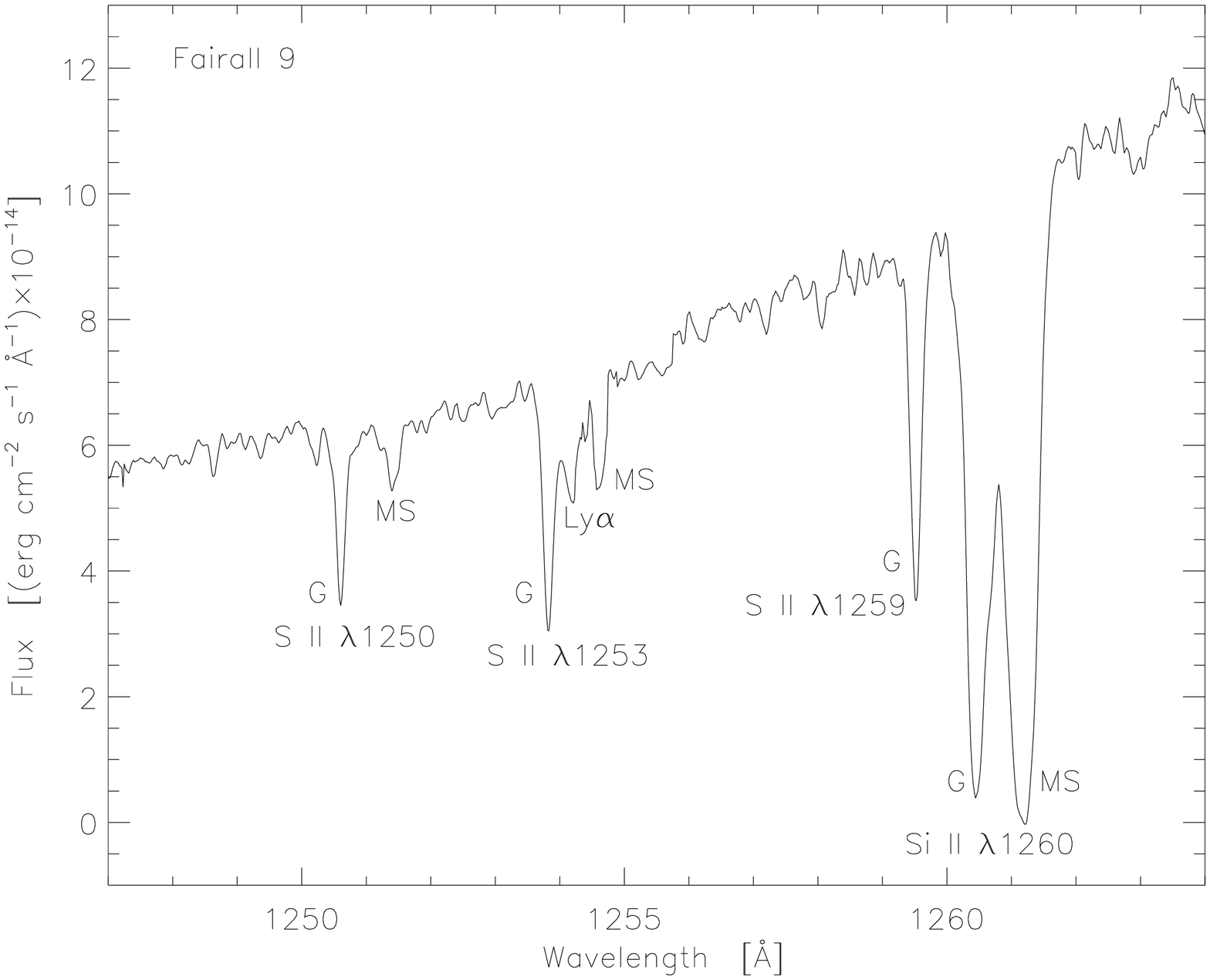}

\clearpage

\epsscale{1.0}
\plotone{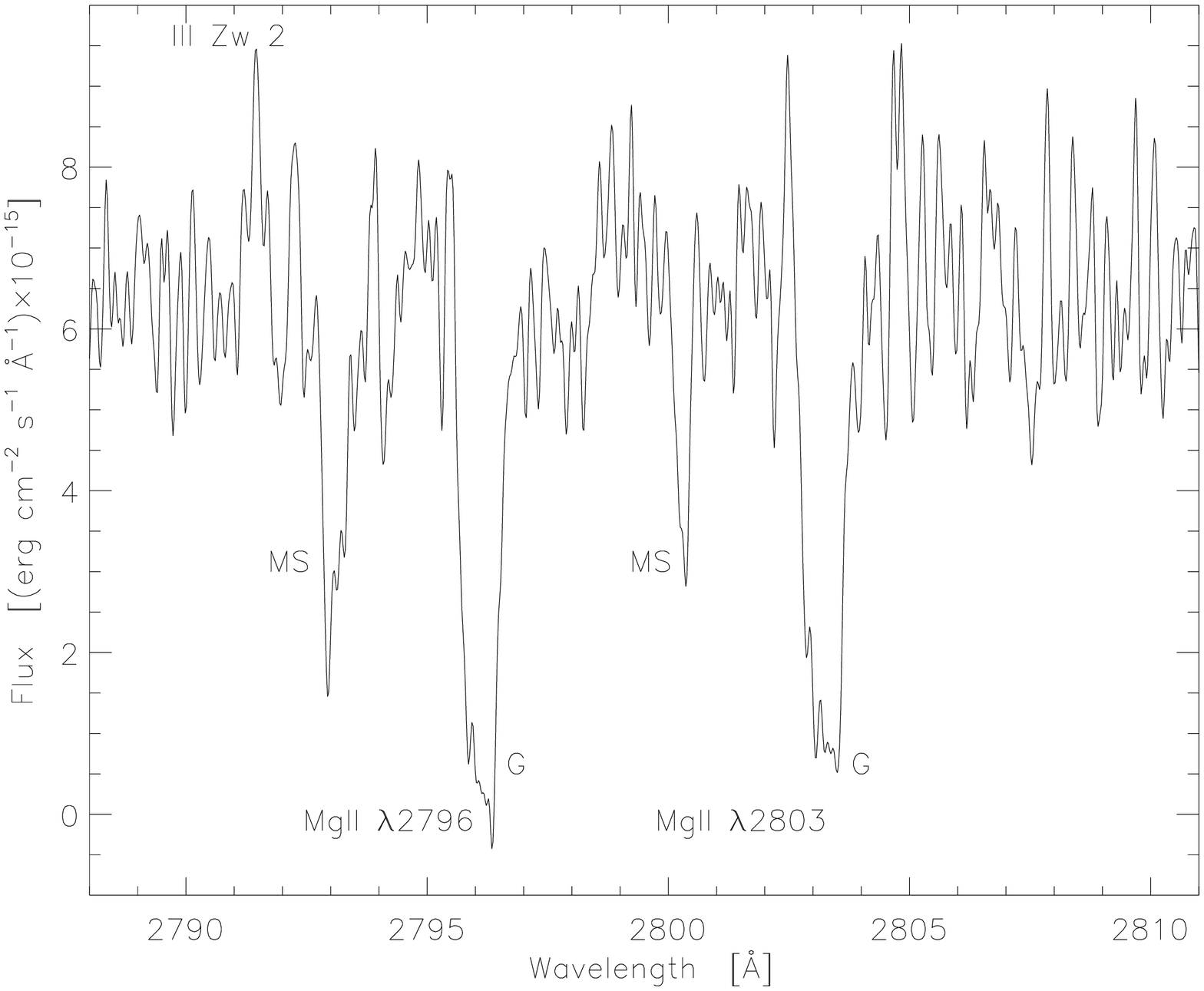}

\clearpage

\epsscale{1.0}
\plotone{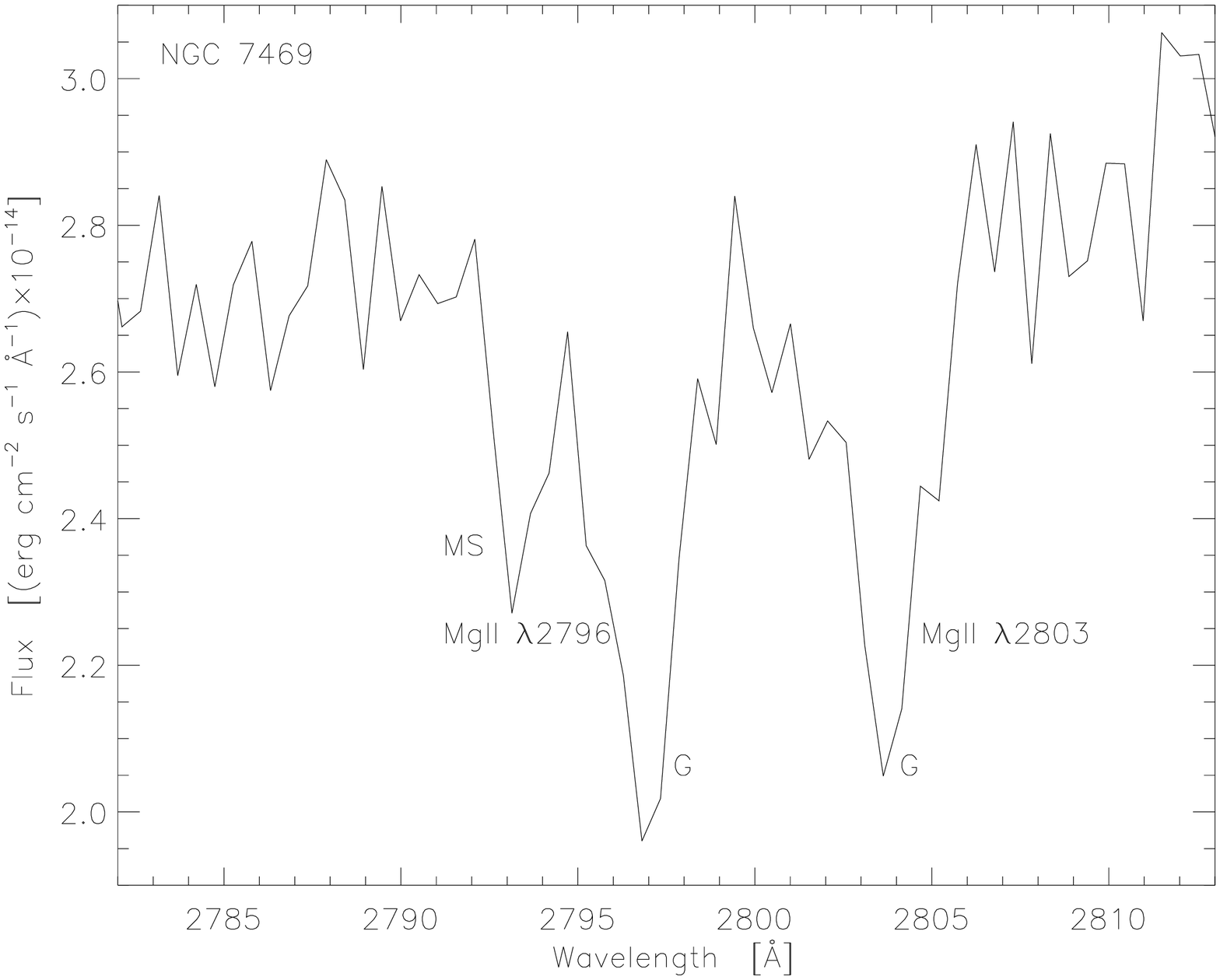}

\clearpage

\epsscale{1.0}
\plotone{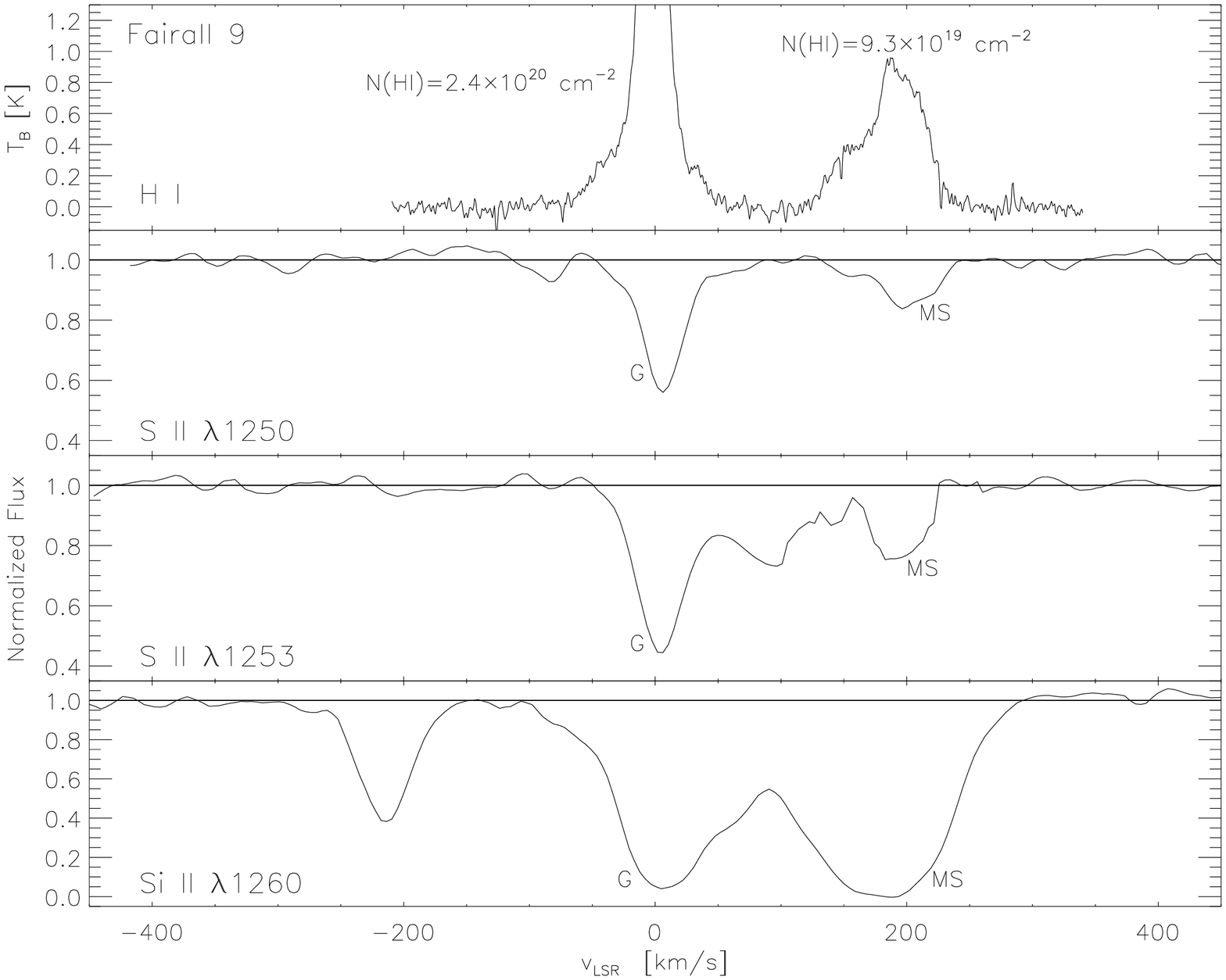}

\clearpage

\epsscale{1.0}
\plotone{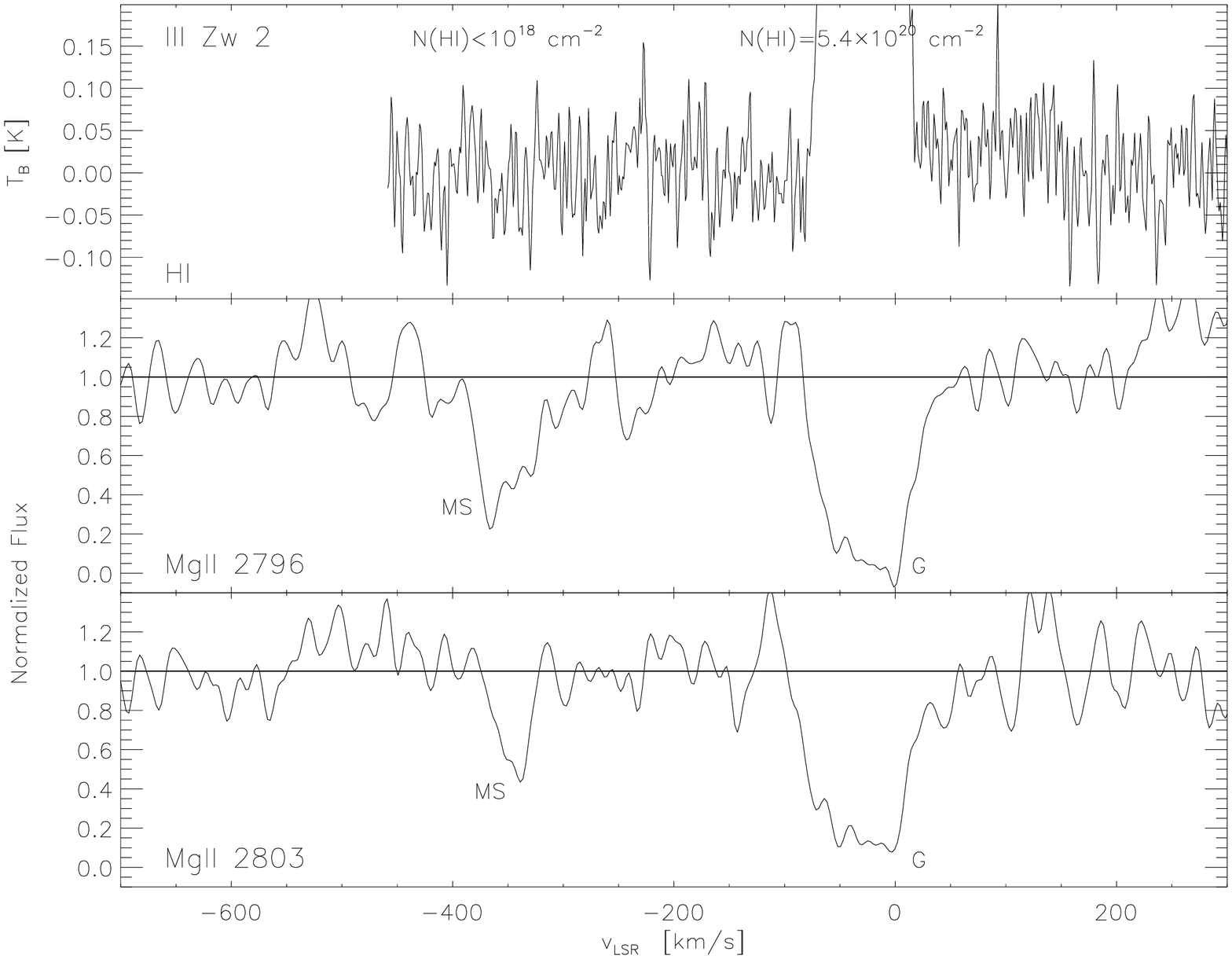}

\clearpage

\epsscale{1.0}
\plotone{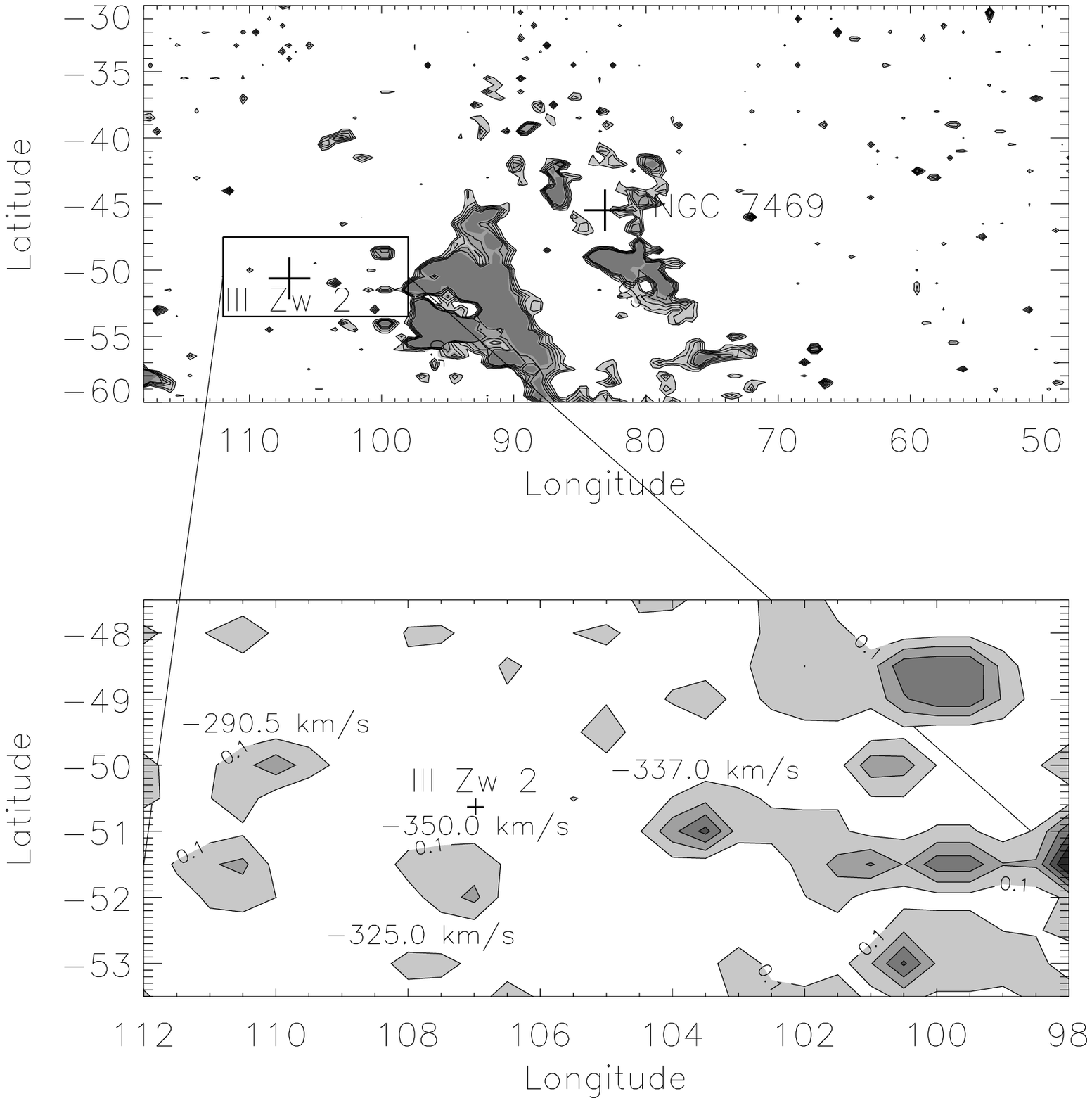}

\clearpage

\epsscale{1.0}
\plotone{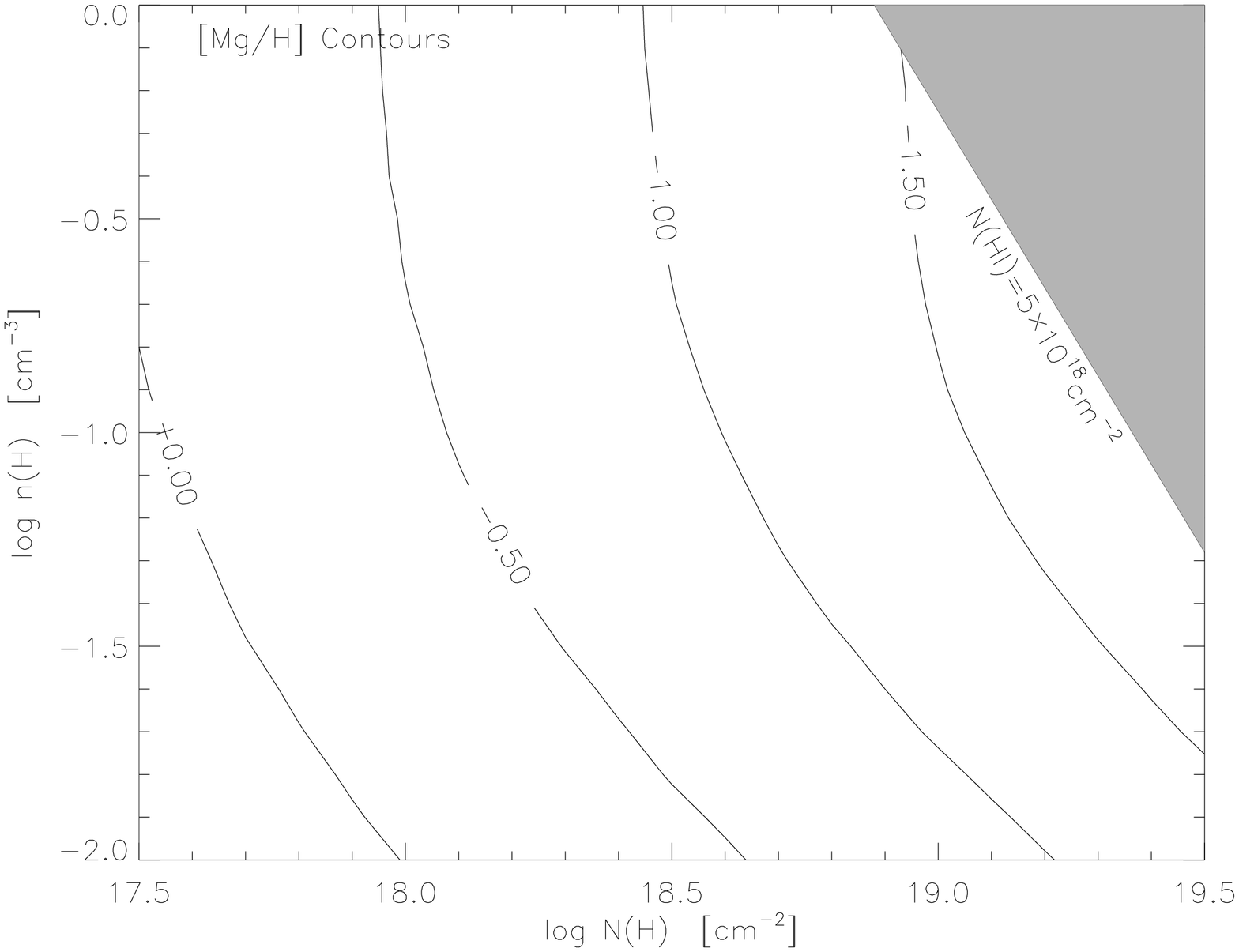}

\clearpage

\epsscale{1.0}
\plotone{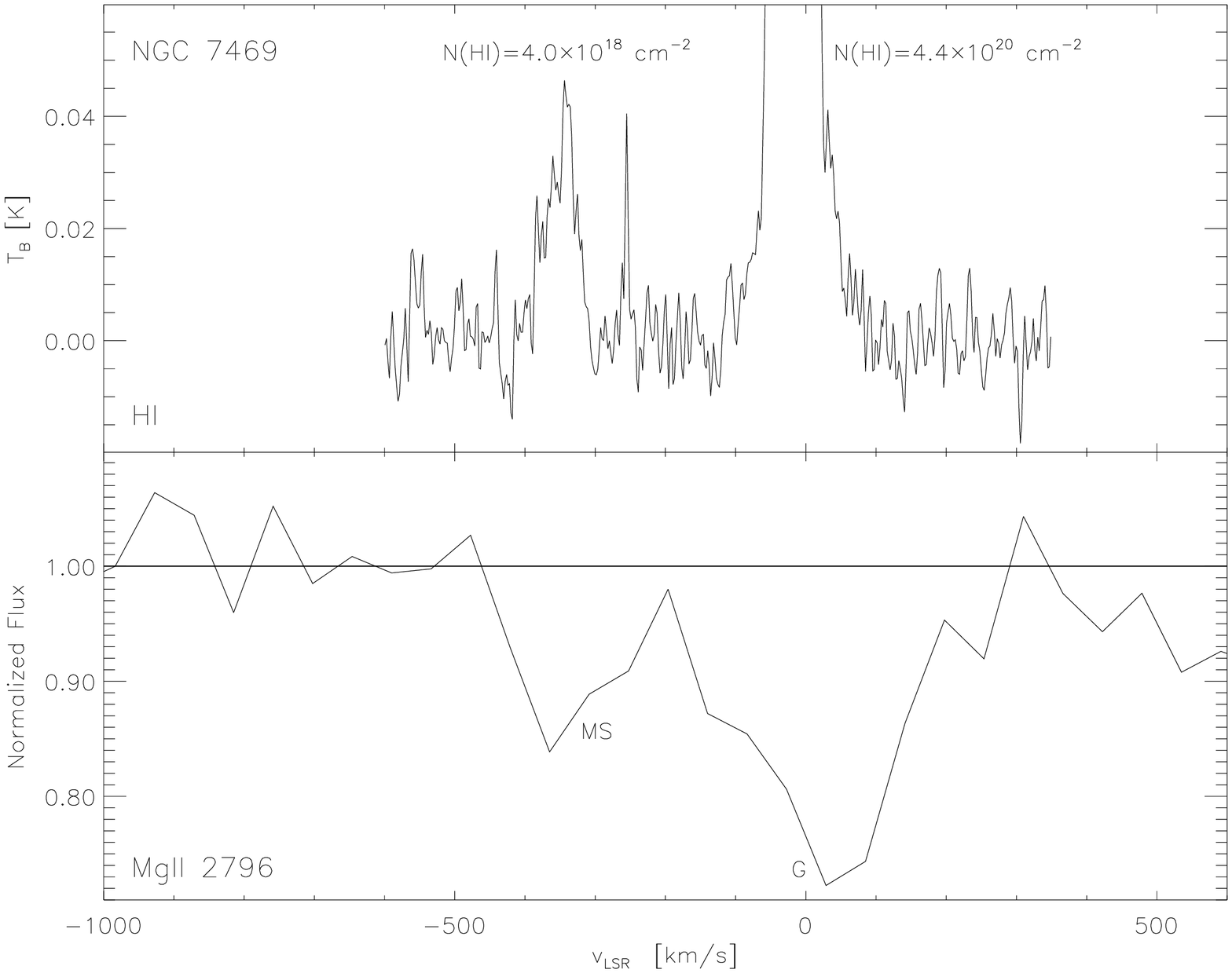}

\end{document}